\documentclass[]{spie}

\usepackage{amsmath,amsfonts,amssymb}
\usepackage{graphicx}
\usepackage[colorlinks=true, allcolors=blue]{hyperref}
\usepackage{comment}
\usepackage{wrapfig}
\usepackage{threeparttable}
\usepackage{floatrow}

\newcommand{\aprx}{\mbox{$\ensuremath{\sim}$}}

\title{Mechanical and Optical Design of the HIRAX Radio Telescope}

\author[a,b]{Benjamin R.~B. Saliwanchik}
\author[c,d]{Aaron Ewall-Wice}
\author[e,f]{Devin Crichton}
\author[a]{Emily R. Kuhn}
\author[g]{Deniz \"{O}l\c{c}ek}
\author[h]{Kevin Bandura}
\author[e,i]{Martin Bucher}
\author[c]{Tzu-Ching Chang}
\author[e,g]{H. Cynthia Chiang}
\author[g]{Kit Gerodias}
\author[e]{Kabelo Kesebonye}
\author[j]{Vincent MacKay}
\author[e]{Kavilan Moodley}
\author[a]{Laura B. Newburgh}
\author[k]{Viraj Nistane}
\author[l]{Jeffrey B. Peterson}
\author[g]{Elizabeth Pieters}
\author[e]{Carla Pieterse}
\author[j]{Keith Vanderlinde}
\author[g,m]{Jonathan L. Sievers}
\author[n]{Amanda Weltman}
\author[g]{Dallas Wulf}

\affil[a]{Department of Physics, Yale University, New Haven, CT, USA}
\affil[b]{Department of Physics, Brookhaven National Laboratory, Upton, NY, USA}
\affil[c]{NASA Jet Propulsion Laboratory, Pasadena, CA, USA}
\affil[d]{Department of Astronomy, University of California, Berkeley, CA, USA}
\affil[e]{School of Mathematics, Statistics, \& Computer Science, University of KwaZulu-Natal, Durban, South Africa}
\affil[f]{Institute for Particle Physics and Astrophysics, ETH Z\"{u}rich, Z\"{u}rich, Switzerland}
\affil[g]{Department of Physics, McGill University, Quebec, Canada}
\affil[h]{Department of Computer Science and Electrical Engineering, and Center for Gravitational Waves and Cosmology, West Virginia University, Morgantown, WV, USA}
\affil[i]{Astroparticle and Cosmology Laboratory, University of Paris, Paris, France}
\affil[j]{Department of Physics, University of Toronto, Toronto, Canada}
\affil[k]{Department of Theoretical Physics and Centre for Astroparticle Physics, Universit\'{e} de Gen\`{e}ve, Gen\`{e}ve, Switzerland}
\affil[l]{Department of Physics, Carnegie Mellon University, Pittsburgh, PA, USA}
\affil[m]{School of Chemistry and Physics, University of KwaZulu-Natal, Durban, South Africa}
\affil[n]{Department of Mathematics, University of Cape Town, Cape Town, South Africa}

\authorinfo{Further author information: (Send correspondence to B.R.B.S.)\\ E-mail: benjamin.saliwanchik@yale.edu}

\begin{document} 
\maketitle

\begin{abstract}

The Hydrogen Intensity and Real-time Analysis eXperiment (HIRAX) is a planned interferometric radio telescope array that will ultimately consist of 1024 close packed 6~m dishes that will be deployed at the SKA South Africa site. HIRAX will survey the majority of the southern sky to measure baryon acoustic oscillations (BAO) using the 21~cm hyperfine transition of neutral hydrogen. It will operate between 400-800 MHz with 391~kHz resolution, corresponding to a redshift range of $0.8 < z < 2.5$ and a minimum $\Delta z/z$ of \aprx 0.003 (frequency resolution $500 < R < 1000$). One of the primary science goals of HIRAX is to constrain the dark energy equation of state by measuring the BAO scale as a function of redshift over a cosmologically significant range. Achieving this goal places stringent requirements on the mechanical and optical design of the HIRAX instrument which are described in this paper. This includes the simulations used to optimize the mechanical and electromagnetic characteristics of the instrument, including the dish focal ratio, receiver support mechanism, and instrument cabling. As a result of these simulations, the dish focal ratio has been reduced to 0.23 to reduce inter-dish crosstalk, the feed support mechanism has been redesigned as a wide (35~cm diam.) central column, and the feed design has been modified to allow the cabling for the receiver to pass directly along the symmetry axis of the feed and dish in order to eliminate beam asymmetries and reduce sidelobe amplitudes. The beams from these full-instrument simulations are also used in an astrophysical m-mode analysis pipeline which is used to evaluate cosmological constraints and determine potential systematic contamination due to physical non-redundancies of the array elements. This end-to-end simulation pipeline was used to inform the dish manufacturing and assembly specifications which will guide the production and construction of the first-stage HIRAX 256-element array. 

\end{abstract}

\section{Introduction}

The fact that the universe is in a state of accelerated expansion today is supported by the observational evidence from various cosmological tools such as the Cosmic Microwave Background (CMB), Type 1a Supernovae (SN1a), and Baryon Acoustic Oscillations (BAO)\cite{wein13}. The increasing rate of expansion is attributed to an unknown cosmological component called Dark Energy, which constitutes around 70\% of the total energy density of the current universe. Current data shows Dark Energy began to affect the dynamics of the universe around a redshift of z $\sim$ 2 (10.5 Gyr ago), and became the dominant component of the energy density at around a redshift of z $\sim$ 0.5 (5.2 Gyr ago). Constraining the properties of Dark Energy, and the evolution of the universe over cosmic timescales, is an important goal of modern cosmology.

Baryon acoustic oscillations provide a standard ruler, a structure of a fixed co-moving scale, which can be used to measure the expansion history of the universe over a wide range of redshifts\cite{Bassett:2009mm}. Baryon acoustic oscillations in the observed matter density power spectrum arise because all modes are initially excited with the same phase on superhorizon scales, and only the growing mode is excited. Before recombination the photons and baryons are tightly coupled to each other owing to Thomson scattering and oscillate much like ordinary sound, following horizon crossing. However, after recombination the photons free stream, causing the phase oscillations of the plasma at the moment of recombination to become frozen into the power spectrum.
The characteristic scale of the first peak of this power spectrum is the sound horizon (comoving distance that density waves in the photon-baryon fluid could have travelled until decoupling), which is strongly constrained by the CMB to be 146.8 $\pm$ 1.8 Mpc \cite{Kogut:2007tq}. Due to the variations in matter density produced by the BAO, and structure formation favoring areas of over-density, the distribution of galaxies subsequently displays a spatial correlation function which corresponds to the BAO power spectrum.
This large scale structure produced by the BAO has been detected at high significance in optical galaxy surveys of the low redshift universe, and has provided an important constraint on cosmological parameters, including Dark Energy\cite{eisenstein05,percival07,abbott18}. 

One method to measure the BAO spectrum at higher redshift is to directly measure the neutral hydrogen density associated with galaxies by using the 21~cm emission line. This probe is particularly attractive because of the omnipresence of neutral Hydrogen at all redshifts. 
Since the characteristic scale of the BAO is very large, there is no need to detect individual galaxies at high resolution, and instead a measurement of the power spectrum of neutral hydrogen density on large scales can completely capture the BAO structure.
This technique is called intensity mapping \cite{Chang:2007xk}.
Galaxies are observed collectively via low spatial resolution measurements of redshifted 21~cm lines of neutral Hydrogen, which also has the potential to make 3D intensity mapping more efficient as compared to optical galaxy surveys. 

The Hydrogen Intensity and Real-time Analysis eXperiment (HIRAX) will measure the HI density field over the redshift range $0.8 < z < 2.5$, bracketing the redshift at which dark energy begins to affect the dynamics of the universe.
The large daily survey area (approximately 1,000 deg$^2$), and real-time processing capabilities of HIRAX will also make it an effective observatory for detecting and monitoring radio transients, such as fast radio bursts (FRBs) and pulsars\cite{chime19a,chime19b,chime19c,chime20a,chime20b}. The Southern Hemisphere survey location also provides overlap with the survey fields of cosmology surveys undertaken by ACT\cite{swetz18}, SPT\cite{benson14}, DESI\cite{levi13}, and the Vera Rubin Observatory\cite{brough20}, allowing for significant cross correlation studies. HIRAX will also deliver a blind HI 21 cm line survey at $0.8 < z < 2.5$.  This will complement the MeerKAT Absorption Line Survey (MALS; $0 < z < 1.4$) by extending the exploration of cold atomic gas in the circum-galactic and inter-galactic medium to higher redshifts\cite{gupta17}.

Detecting the BAO in the presence of significant astronomical foregrounds, most notably synchrotron radiation from the Milky Way galaxy, is a significant challenge. The high level requirements for the array to achieve its science goals are sensitivity, redundancy, and control of systematics.
The array is designed to meet the sensitivity requirements by a combination of low system temperature and large collecting area. The later aspect is what drives us to a large array, consisting of approximately one thousand 6~m dishes. The former is achieved by developing low loss antennas and amplifiers, which are addressed in Kuhn et al.\cite{kuhn20}. The simulations presented in this work are part of the effort to design and produce an array with very high levels of redundancy between elements, and with very low systematic levels.
  
Tight control over redundancy is a new design driver that has emerged for the next generation of 21~cm observatories such as HIRAX\cite{newburgh16}. We are aiming to solve the problem of redundancy in hardware, which requires a level of precision that is significantly higher than existing experiments such as CHIME\cite{bandura14} and HERA\cite{deboer17}. 
That is, rather than correcting for the differences between array element structures and positions completely in software, we aim to reduce the element differences as much as possible in hardware, before applying fine corrections in software. 
This requires significantly more stringent specifications than normal for a radio telescope, on the order of 1 part in 1000 relative to the wavelength, driven largely by the high foreground to signal ratio for the 21~cm signal. Edge effects due to the finite size of the array are also a source of non-redundancy, and must be taken into consideration. Additionally, we are also concerned with minimizing crosstalk between the elements in the array, which can result from either sky signals bouncing between elements, or from amplifier noise being transmitted from one element and picked up in another. The issues we want to address in this work are, broadly, how to optimize the design of the instrument to minimize systematics, and to determine what level of element redundancy is necessary, and whether this level is feasible to achieve in hardware, within cost.

\section{Telescope Design}

\begin{figure}[H]
\begin{center}
\includegraphics[width=0.8 \textwidth]{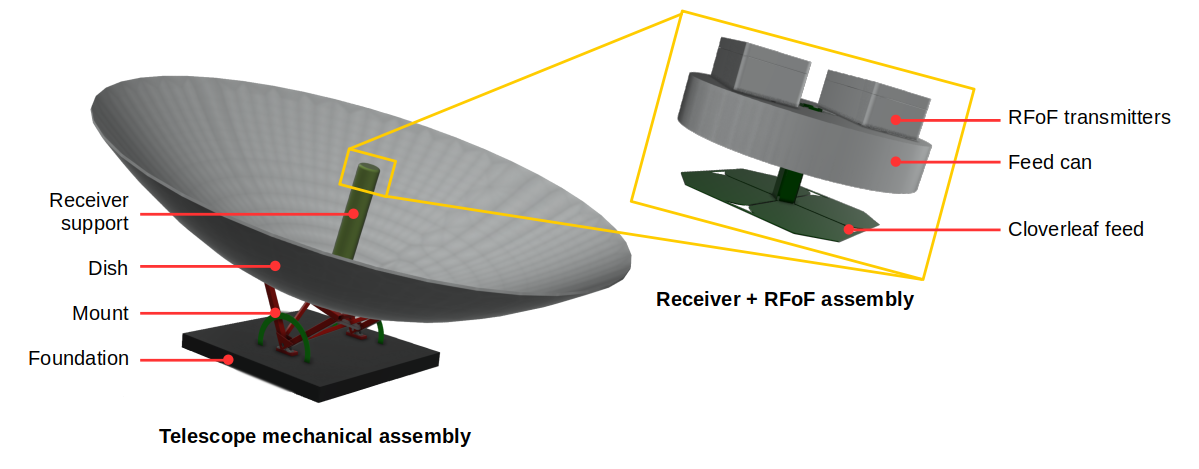}
\includegraphics[width=0.8 \textwidth]{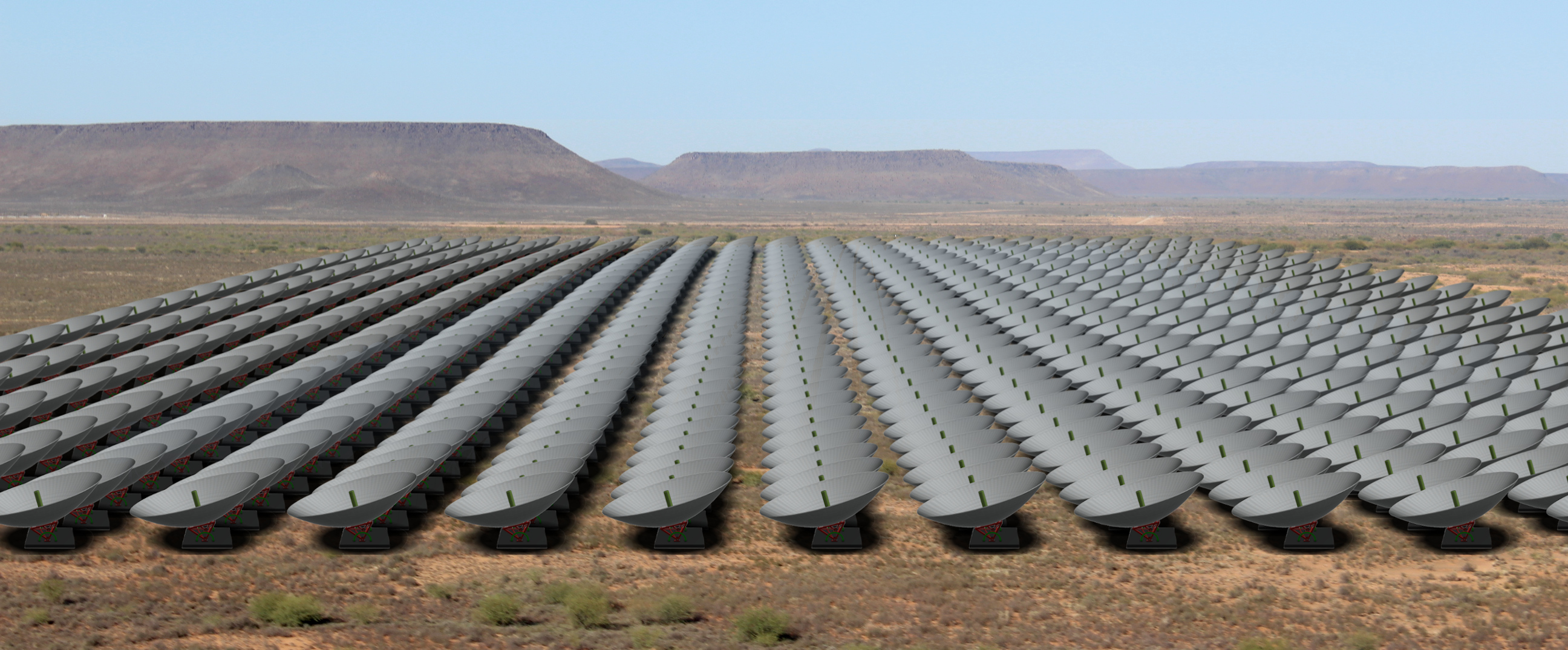}
\end{center}
\caption[]{Top: Conceptual illustration of the HIRAX telescope design. The dishes are 6~m in diameter, with a focal ratio of 0.23.
The receiver is supported by a fiberglass column which allows for axisymmetric cabling, which is important for beam symmetry, polarization, and low sidelobe amplitudes, and which is rigid under high winds. The receiver is a dual-polarization cloverleaf antenna, with a metal ``can'' structure to provide a backplane, and reduce spillover and crosstalk. Analog signals from the dishes are transmitted via RF over fiber to the back end correlator. Bottom: A rendering of the full 1024 element array. The array is close-packed to provide improved sensitivity on the BAO angular scales, and to provide highly redundant baselines, which facilitate calibration and correlation.}
\label{fig:dish_array_model} 
\end{figure}

The full HIRAX array will consist of 1024 parabolic dishes 6~m in diameter, with a focal ratio of f/D = 0.23. 
The optimization of the focal ratio is discussed in Section \ref{sec:design_sims}. A first-stage array consisting of 256 elements is fully funded, and bidding for construction of the array is underway. The full array will be arranged in a close-packed $32 \times 32$ element configuration to enhance sensitivity on the BAO length scales, and to increase the redundancy of the array. The increased array redundancy in turn simplifies calibration, reduces the number of correlations to be calculated, and reduces the data volume to be stored. The telescope design and array layout can be seen in Figure \ref{fig:dish_array_model}.

The dish design includes a low mount that allows for easier access to the feed, reduces wind force on the base, and reduces cost. The HIRAX array operates in drift scan, so the only degree of freedom necessary in the dish pointing is to vary the elevation between $\pm30^\circ$ degrees from zenith to encompass the intended survey field.

The receiver is supported by a central fiberglass column.  The architecture of this support structure was motivated by the optimal path of the cables from the feed to the surface of the dish.  Because the cables are metal and lie in the optical path, the sidelobes of the primary beam depend sensitively upon the cable placement and angle.  Simulations show that sidelobe levels and asymmetry are minimized when the cables run straight down the boresight axis of the dish, rather than at an inclined angle, as discussed in Section \ref{sec:design_sims}. A column was therefore chosen to provide a natural environmental enclosure following this cable path. The support structure includes additional provision for fully enclosing the feed and the radio-frequency over fiber (RFoF) modules that are co-located with the feed, for weatherproofing and protection of the full receiver system.

The HIRAX feed is a dual-polarization cloverleaf antenna consisting of copper layers in an FR-4 (printed circuit board) composite (shown in Figure \ref{fig:cst_model}). This method of manufacturing is easy to produce at scale, easy to assemble with a minimal jig for alignment, and the FR-4 protects the metallic elements from corrosion. The cloverleaf antenna is a proven design, having been deployed on the Canadian Hydrogen Intensity Mapping Experiment (CHIME) \cite{deng14}. While the original CHIME feeds were passive, the versions developed for HIRAX integrate the first stage low noise amplifier (LNA) with the antenna balun to reduce the system noise. As a result of the lower system noise, the feed material was changed to FR-4 (compared to teflon in CHIME), which has higher loss, but is substantially cheaper. In an additional change from the CHIME architechture, the signal from the individual array elements is transmitted to the correlator by RFoF, instead of via coaxial cable.
This reduces cost relative to coax for a large array, with no loss to performance. Noise temperature measurements of the HIRAX feeds are being conducted using a cryogenic RF chamber to perform a differential y-factor measurement between hot (295~K) and cold (77~K) loads. Details of those measurements are presented in Kuhn et al. \cite{kuhn20}.

Since we intend to use our EM simulations to optimize elements of the feed and dish design, we wanted to verify that those simulations were accurate. We conducted the first range measurements of the HIRAX feed beams at the MESA antenna test facility at the NASA Jet Propulsion Laboratory, and further measurements at North Carolina State University. Figure \ref{fig:nc_range} shows the measured beams of the HIRAX cloverleaf antenna and can, and compares this to the CST simulations. These measurements confirm the accuracy of the simulations, and that the feeds were produced according to design.

\begin{figure}[H]
\floatbox[{\capbeside\thisfloatsetup{capbesideposition={right,center},capbesidewidth=0.3 \textwidth}}]{figure}[\FBwidth]
{\includegraphics[width=0.450 \textwidth]{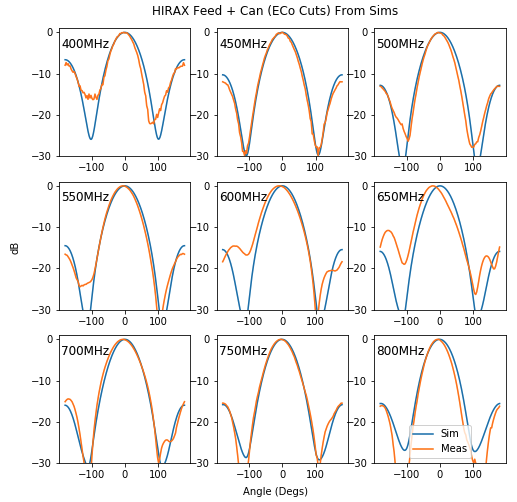}}
{\caption{E-plane co-polarization beam measurements of the HIRAX cloverleaf antenna and can (orange), compared to the simulated beams (blue) in 50~MHz increments across the HIRAX observing band. These measurements confirm the beam width and gain are as designed and simulated. Differences in the depth of nulls are due to low amplitude reflections in the range, which are not present in free-space simulations. Measurements were taken at the North Carolina State University test range.}\label{fig:nc_range}}
\end{figure}

Work is currently underway to develop a drone calibration platform, which will allow for beam measurements of the full instrument, including the 6m dish and amplification chains, in order to verify the simulated performance of the instrument, fully characterize the instrument, and aid in calibration. This platform has already been used to perform beam measurements of the Baryon Mapping Experiment (BMX) array at Brookhaven National Laboratory\cite{oconnor20}. Beam measurements of the HIRAX dishes is currently being delayed due to the ongoing Covid-19 pandemic, but will be performed when international research travel is permitted by institutions again.

\section{Electromagnetic Simulations}

\subsection{Design Simulations}
\label{sec:design_sims}

Electromagnetic simulations of the instrument were performed to select between design options, and to optimize the design elements. Instrument elements explored include:

\begin{itemize}
    \item The diameter of the feed ``can'' structure    
    \item The nature and positioning of the power cabling for the feed
    \item The feed mechanical support mechanism, in particular deciding between a system with several ``feed legs'' and a monolithic ``feed column''
    \item Optimizing the dish focal ratio to reduce crosstalk and other array effects    
\end{itemize}

\begin{figure}[H]
\begin{center}
\includegraphics[width=0.4 \textwidth]{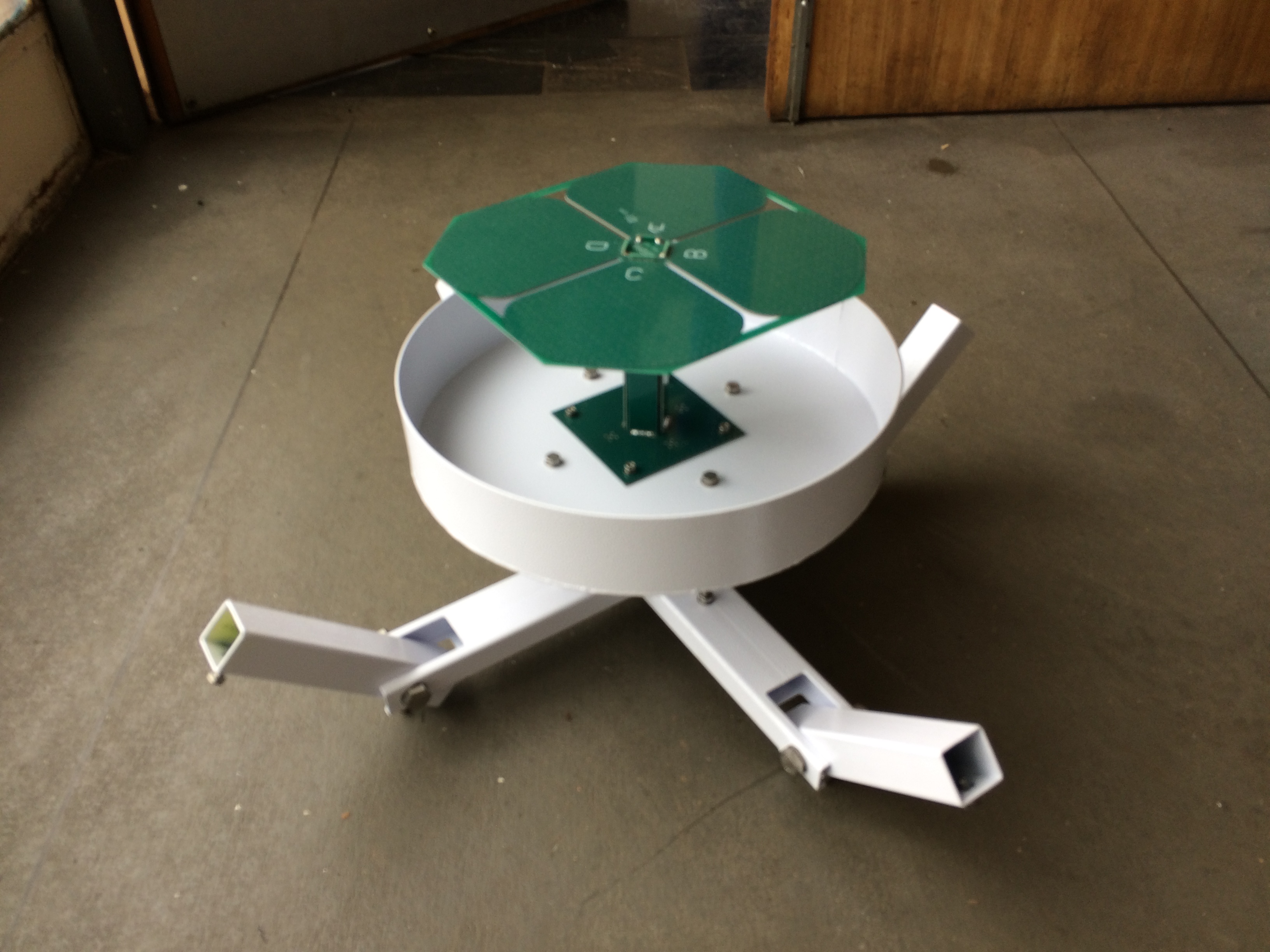}
\includegraphics[width=0.4 \textwidth]{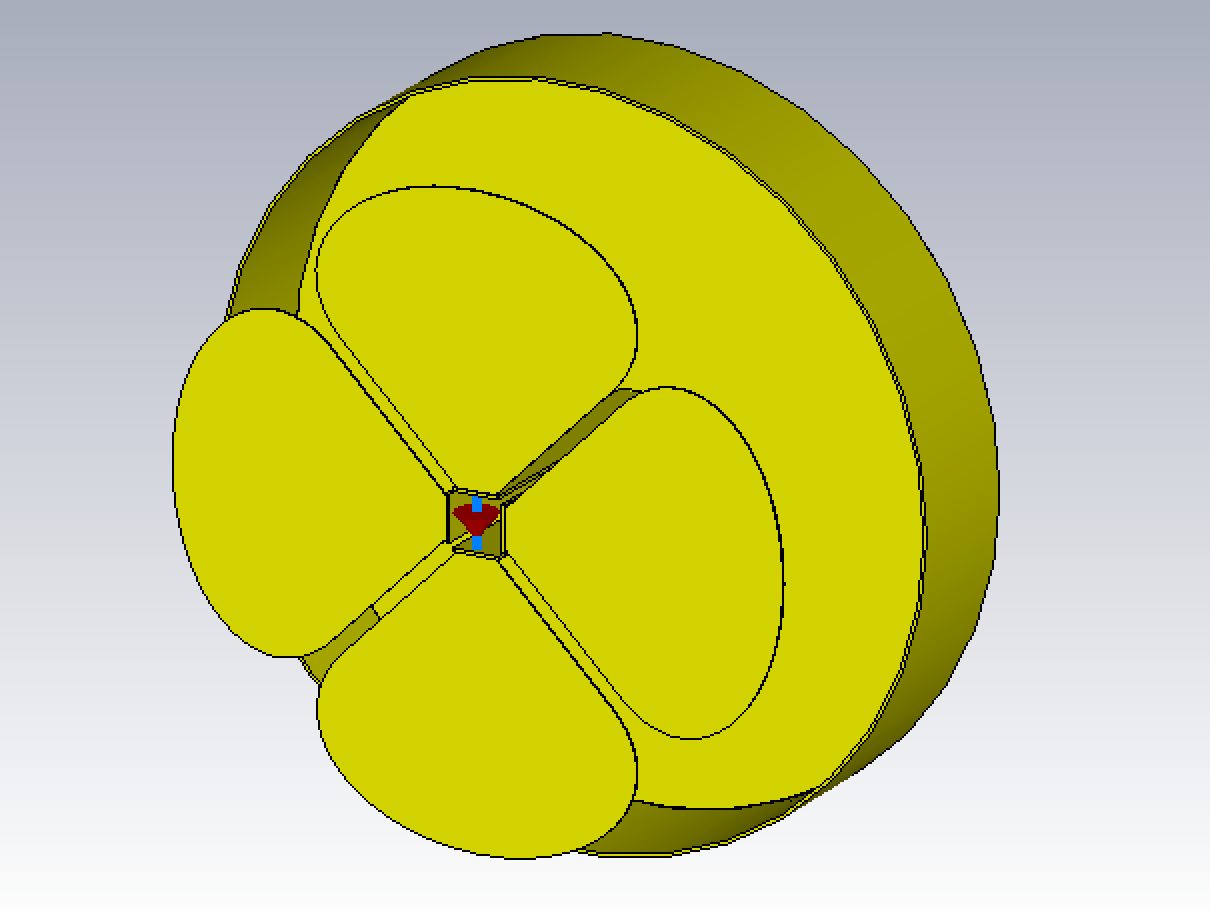}
\includegraphics[width=0.4 \textwidth]{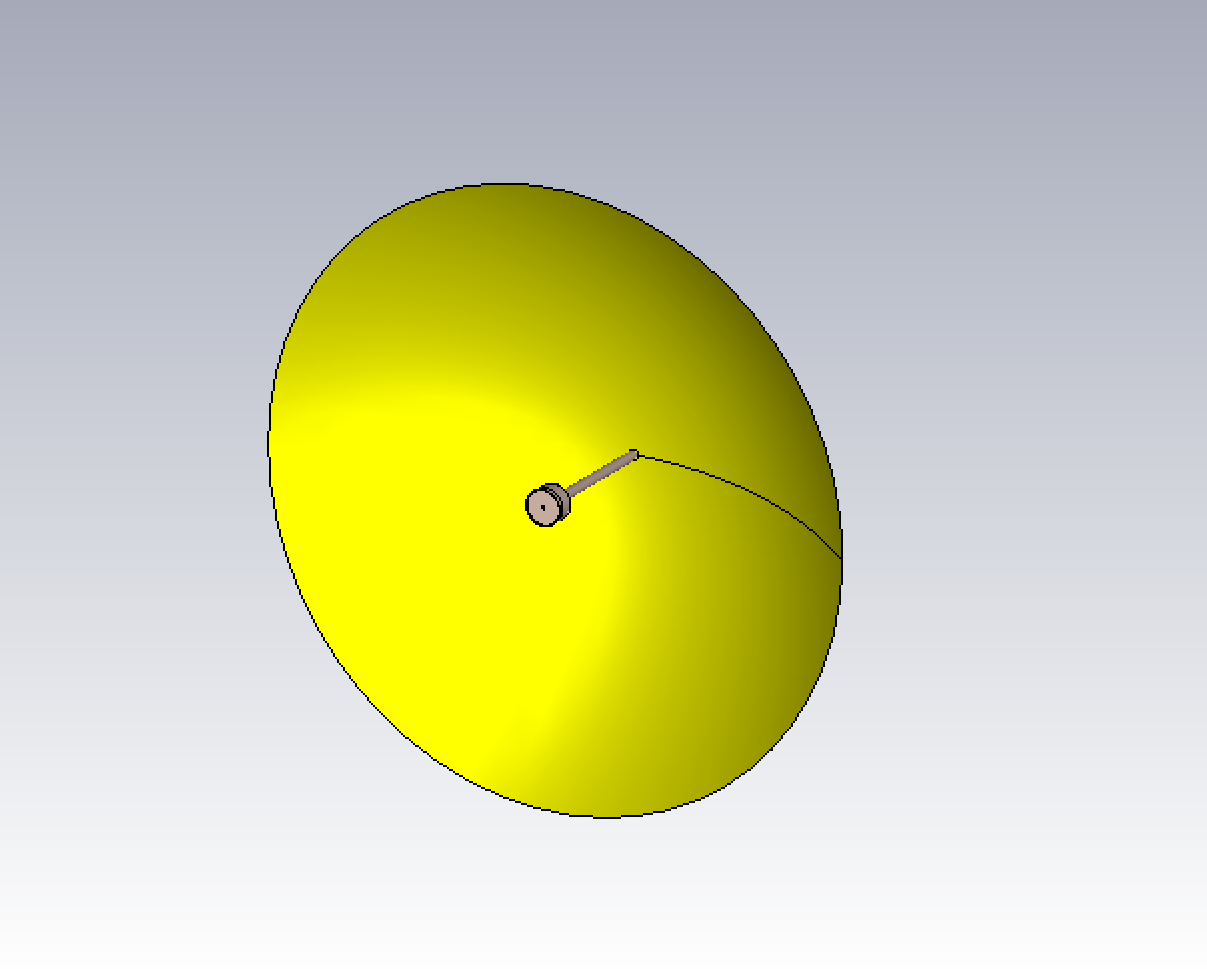}
\includegraphics[width=0.4 \textwidth]{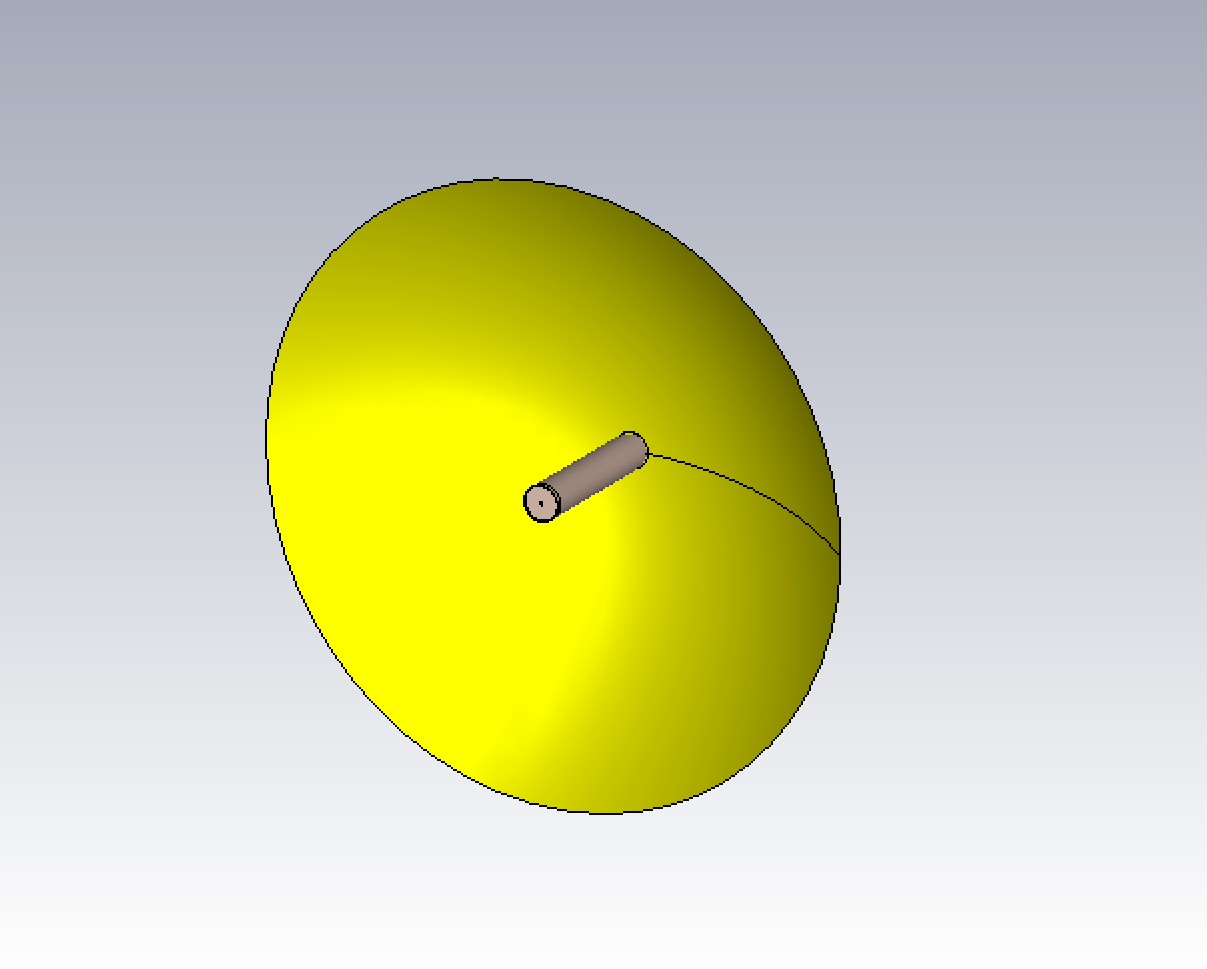}
\end{center}
\caption[]{The HIRAX feed and CST models. Top left: The HIRAX dual-polarization cloverleaf antenna (green PCB) and can (white). Top right: CST model of the HIRAX feed and can. Only one polarization is fed in the simulation, as indicated by the blue bar and red arrow. Bottom left: The 6~m dish and 10~cm diameter fiberglass feed column model. Bottom right: The dish and 35~cm diameter feed column model.}
\label{fig:cst_model} 
\end{figure}

\subsubsection{Feed Can} The HIRAX cloverleaf antenna is backed with a metal structure referred to as the ``can'' (Figure \ref{fig:cst_model}), a combination of a ground plane and a cylindrical surface which helps to circularize the beam and reduce the beam FWHM, to prevent over illuminating the dish, and reduce spillover. Spillover is especially a concern in a close-packed array, as proximity of the neighboring dishes increased pickup of spilled power. Without the can, the FWHM of the cloverleaf alone is $90^\circ$ ($60^\circ$)  at 400~MHz (800~MHz), while with a nominal 330~mm diameter can the FWHM is reduced to $70^\circ$ ($67^\circ$) at 400~MHz (800~MHz). Increasing the can diameter reduces $S_{11}$ because it provides a better ground plane, but it also decreases aperture efficiency, because it blocks more of the beam reflected from the dish. The effect on $S_{11}$ was found to be a slow function of radius, while the loss of aperture efficiency scales roughly with $r^2$. Therefore a small diameter can was used; $330$~mm in diameter, slightly larger than the cloverleaf itself.

\subsubsection{Feed Support} In initial prototype dishes the feed was supported above the dish by four cylindrical aluminum conduit ``legs'' 25~mm in diameter, forming the edges of a square right pyramid, with the legs meeting at an apex behind the feed, and an apex angle of approximately $45^\circ$. However, simulations showed that scattering off the metal feed legs introduced strongly polarized components into the beam, and increased sidelobe levels, motivating a design change to non-conductive support structure.

Similar feed legs could have been constructed out of a non-conductive composite material such as fiberglass, but there were additional mechanical problems with the feed leg model. Positioning the feed could be accomplished by a mechanism that slid over the conduit legs, and could adjust the position of the receiver independently on each leg. However, this mechanism did not translate easily into adjustments in a simple orthogonal coordinate system. Furthermore, it was prone to motion in several axes, including compression towards the dish (z-axis), and rotation around the z-axis. Weatherproofing and feed access were also issues with the feed leg support mechanism.

A proposed solution was to instead use a fiberglass column, extending from the vertex of the dish to the feed (Figure \ref{fig:cst_model}). Mechanical FEA simulations showed this feed support mechanism simultaneously solved several problems. It was significantly more stable in all axes, particularly the z-axis, which is especially important for ensuring the instrument is in focus. Large diameter cylinders are especially strong, and a 35~cm diameter cylinder, just larger than the feed can, was found to be able to withstand the peak survival wind speed of 44.4~m/s at the SKA South Africa site, while deflecting from the focal point by less than 1~mm. A feed column also provided a weather proof enclosure, easy access to the feed and RFoF module by a detachable section at the top of the cylinder, and a stable point from which fine adjustments in feed position could be made in an orthogonal basis system. A similar system was employed by the CBASS collaboration, using an RF transparent receiver support column constructed of Plastazote LD45, a closed-cell polyethylene foam\cite{king10}.

\begin{figure}[H]
\floatbox[{\capbeside\thisfloatsetup{capbesideposition={right,center},capbesidewidth=0.35 \textwidth}}]{figure}[\FBwidth]
{\includegraphics[width=0.6 \textwidth]{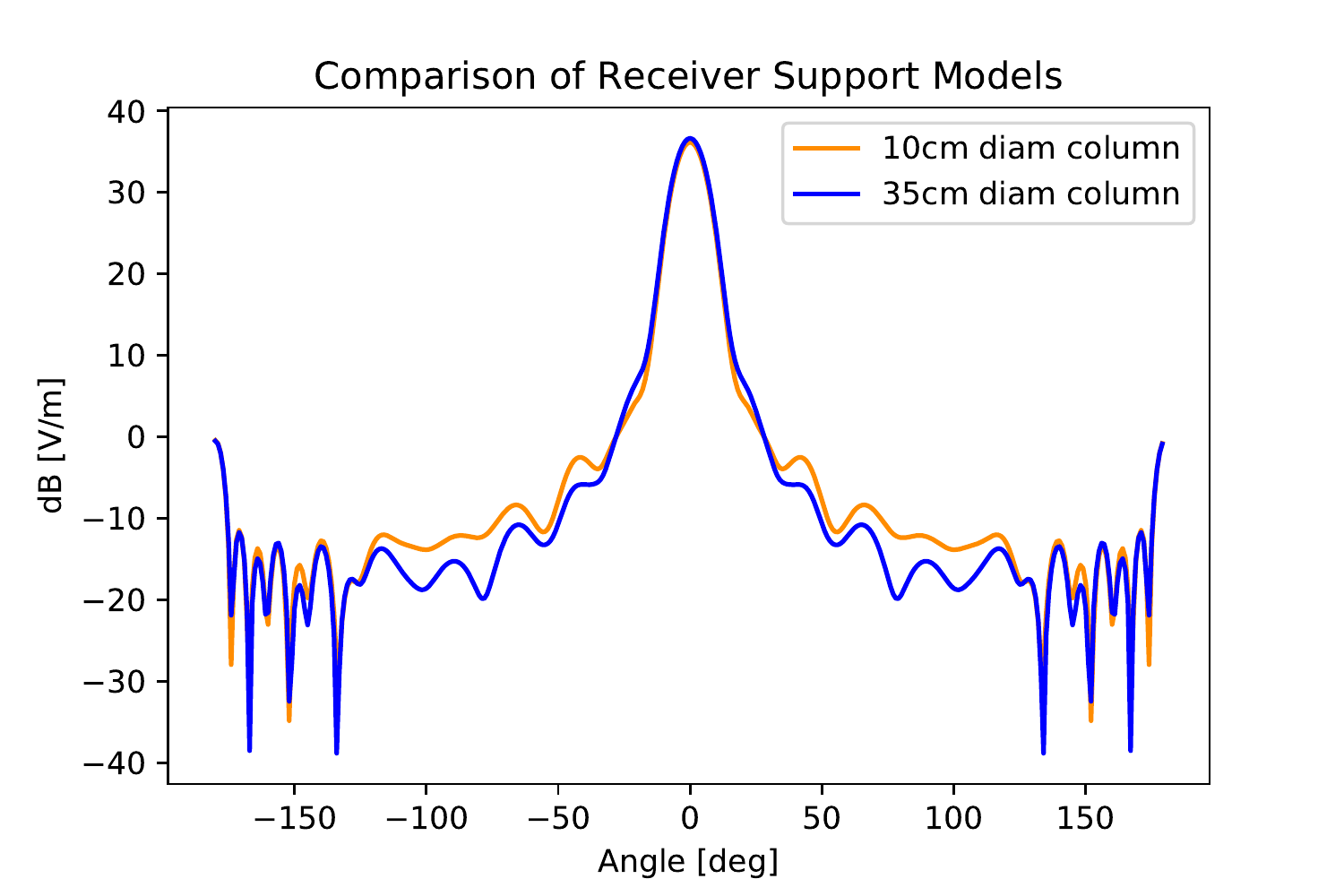}}
{\caption{Comparison of 10~cm diameter receiver support column (orange) and 35~cm column (blue), with 5~mm wall thickness, at 400~MHz. A smaller diameter support column slightly decreases main beam amplitude (by 0.5~dB). Sidelobes can be slightly decreased or enhanced, depending on frequency.} \label{fig:receiver_support}}
\end{figure}

A small diameter (10~cm) feed column was also proposed, and a prototype was fielded in the dish shown in Figure \ref{fig:dish}. However, in addition to greater mechanical stability for fixed wall thickness, a large diameter column also has some RF advantages. EM simulations showed that a small diameter column reduces the main beam amplitude by up to 0.5~dB, due to the long path length of dielectric material along the main beam line of sight (see Figure \ref{fig:receiver_support}). Both models slightly alter the sidelobe amplitudes, but in different directions at different frequencies. A larger diameter column does slightly reduce the instrument radiative efficiency compared to a smaller diameter column, for fixed wall thickness, because of the greater net amount of dielectric in the beam path, but this effect is also not significant. There was a $<1\%$ difference in radiative efficiency between 10~cm and 35~cm diameter columns with 5mm wall thickness, several times thicker than what is mechanically required.

\subsubsection{Power Cabling} In connection with the issue of feed support, it was suspected that the cables for powering the LNA in the feed might affect the instrument beams. Simulations showed that a single coax cable running from the feed to the dish edge induced polarization structures in the beam and enhanced sidelobe structure at a level comparable to the feed support legs themselves, because an incident plane wave polarized in the same direction as the cable will excite currents along it. Additionally, if there is only one cable, and not a symmetric set of feed legs, the sidelobe effect is asymmetric, which is undesirable. 

This effect can be reduced, but not eliminated, by reducing the cable diameter. Significant improvements were seen by reducing from coax ($\aprx1$~cm diam.) to 15 AWG wire ($\aprx1.5$~mm diam.). Reductions below approximately 1~mm diameter produce diminishing returns. 15 AWG wire is sufficient for the current required to power the LNA (200mA), and since in the full array the sky signal will be transmitted on RFoF, neither coax, nor any other metal cable elements will be required to run from the dish to the feed, further reducing scattering.

\begin{figure}[H]
\begin{center}
\includegraphics[width=0.75 \textwidth]{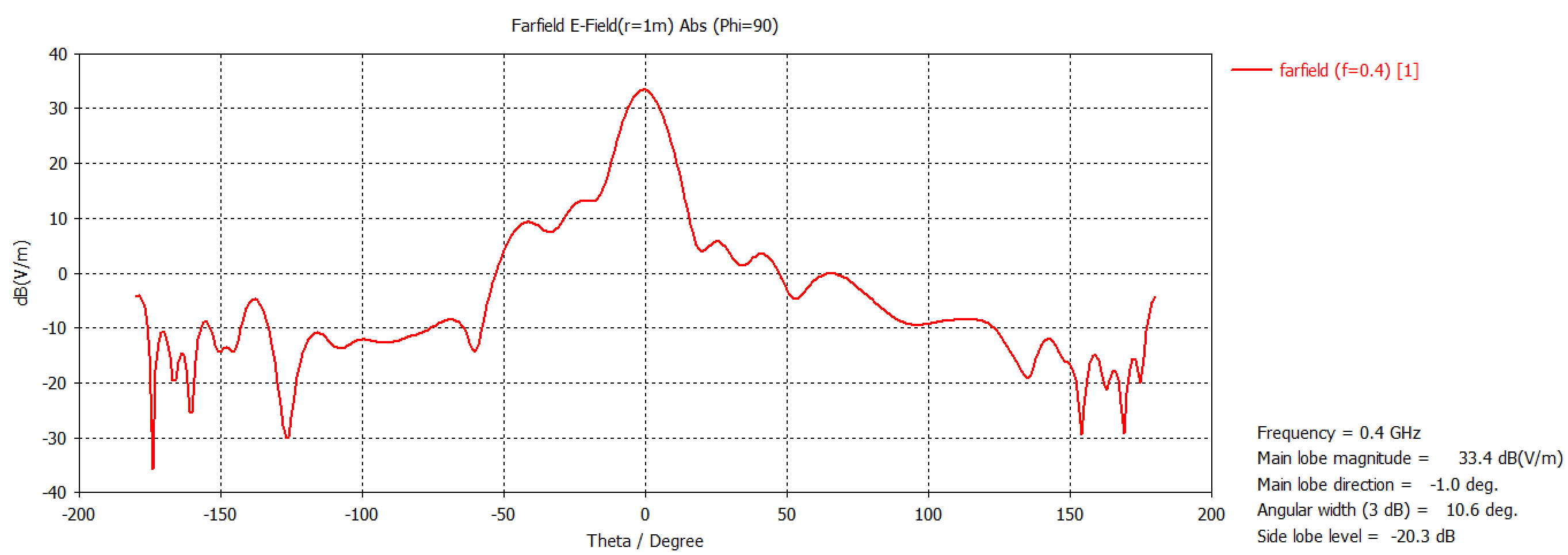}
\includegraphics[width=0.75 \textwidth]{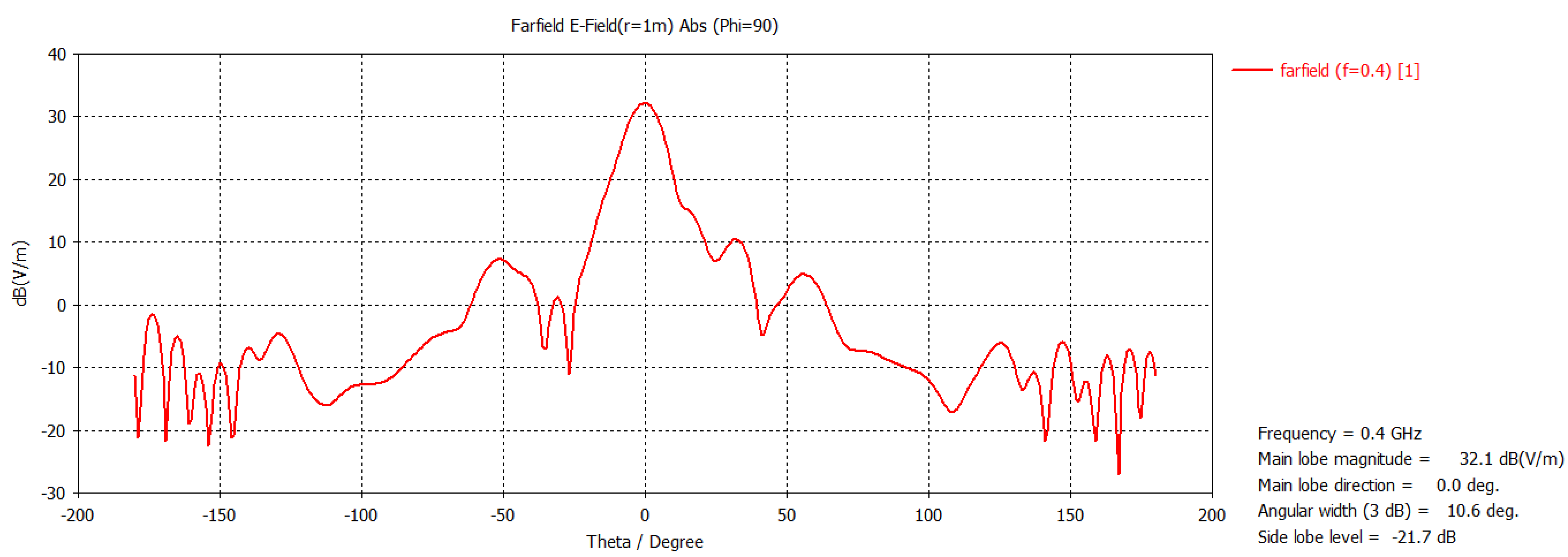}
\includegraphics[width=0.75 \textwidth]{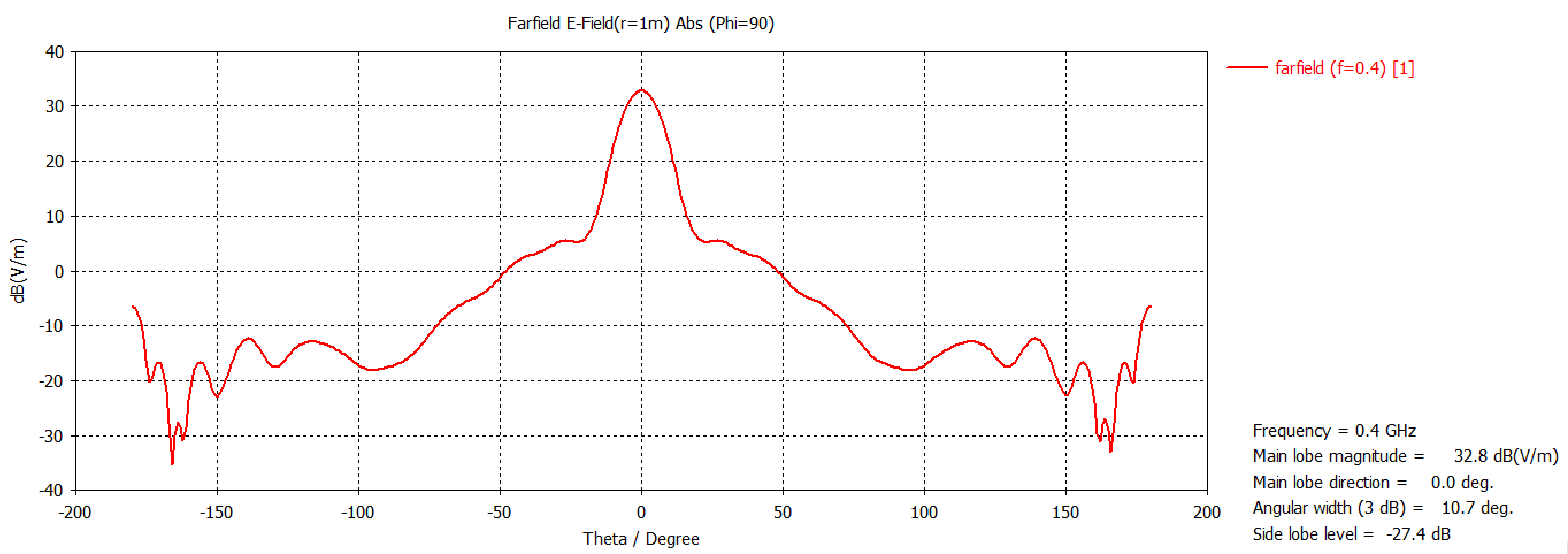}
\end{center}
\caption[]{HIRAX telescope beams at 400~MHz using 15 AWG wire ($\aprx 1.5$~mm diameter) to power the LNA embedded in the balun of the feed. Top: wire routed from focal point to edge of dish. Middle: wire routed from the edge of the feed can ($\aprx18$~cm offset from the axis of symmetry of the dish and antenna) to the vertex of dish. Bottom: wire routed though cloverleaf balun, precisely along the boresight axis of the antenna, to the dish vertex. The completely symmetric wiring case eliminates all asymmetric beam features, and reduces the sidelobe amplitudes by up to 7~dB relative to the initial configuration.}
\label{fig:feed_cabling} 
\end{figure}

The asymmetric sidelobes from the cabling can be further reduced by orienting the power cables parallel to the beam, that is, running them from the feed to the dish vertex. Simulations were run exploring a range of cable angles, between 90 degrees (perpendicular to the beam axis) and zero degrees (parallel to the beam axis), and with the feed side of the cable positioned at a radius from the feed center greater than the can radius (because it mechanically had to wrap around the can to access the feed). The sidelobe behavior was not a monotonic function of angle, but the asymmetry was generally reduced with increasing angle. However, even at zero degrees, the effect was not completely eliminated, as there is still a small component of the cable perpendicular to the beam, due to the offset from the feed center. Running the cable down the exact symmetry axis of the feed and telescope, however, does completely remove the asymmetric sidelobe effect (See Figure \ref{fig:feed_cabling}). This configuration also reduces the sidelobe amplitude by up to 7~dB relative to the initial (perpendicular) wiring case.

We determined that the feed could easily be modified to allow for this cable routing since the balun is already hollow, and a hole in the base board could be added to the PCB design to allow the cable to pass through. There was a minor concern that this would require slightly offsetting the feed point of the antenna, which has also been shown to produce beam asymmetry. However, simulations showed that as long as both the cable and the feed point were both within approximately the balun diameter ($\aprx1$~cm), which is mechanically feasible, then the effects of both offsets are negligible. The feed design has therefore been modified to allow the power cable to be exactly on axis, and the feed point to be $< 1$~cm offset from center.

\subsubsection{Dish Focal Ratio} Reflections within and between an interferometer's antenna elements introduce spectral structure in the antenna gain at frequency scales corresponding to the time delay of the reflection. If these reflections reach to fine frequency scales, they can cause the otherwise smooth continuum foregrounds to contaminate the spectral scales important for cosmology. Lowering the focal ratio reduces cross coupling between feeds, reducing reflections between elements but also has the potential to increase reflections within a single antenna element.

We evaluated the relative delay performance of dishes with different focal ratios, by running CST simulations of a plane-wave excitation of the dish and feed from zenith, and use the voltages in a 50\,$\Omega$ termination of the feed to estimate the time-domain response kernel that leaks these foregrounds using the formalism from Ewall-Wice et al.\cite{EwallWice:2016}. In order to evaluate the relative contributions from inter-dish versus intra-dish reflections, we perform our simulations with two different sets of boundary conditions. First, we set our boundary conditions to be open, simulating a dish in isolation, which gives us the contributions from reflections within the dish. Second, we perform simulations with periodic rectangular boundary conditions to simulate one of the HIRAX antennas embedded within an infinite rectangular packed array. This response function corresponds to the spectral structure that would appear in an autocorrelation for a sky with only a source at zenith. While auto-correlations are not very sensitive to 21\,cm fluctuations, the simulations are useful for evaluating the relative levels of spectral structure. The impact of intra-dish reflections on cross-correlations between dishes is the subject of ongoing work.

\begin{figure}[H]
    \centering
    \includegraphics[width=.8\textwidth]{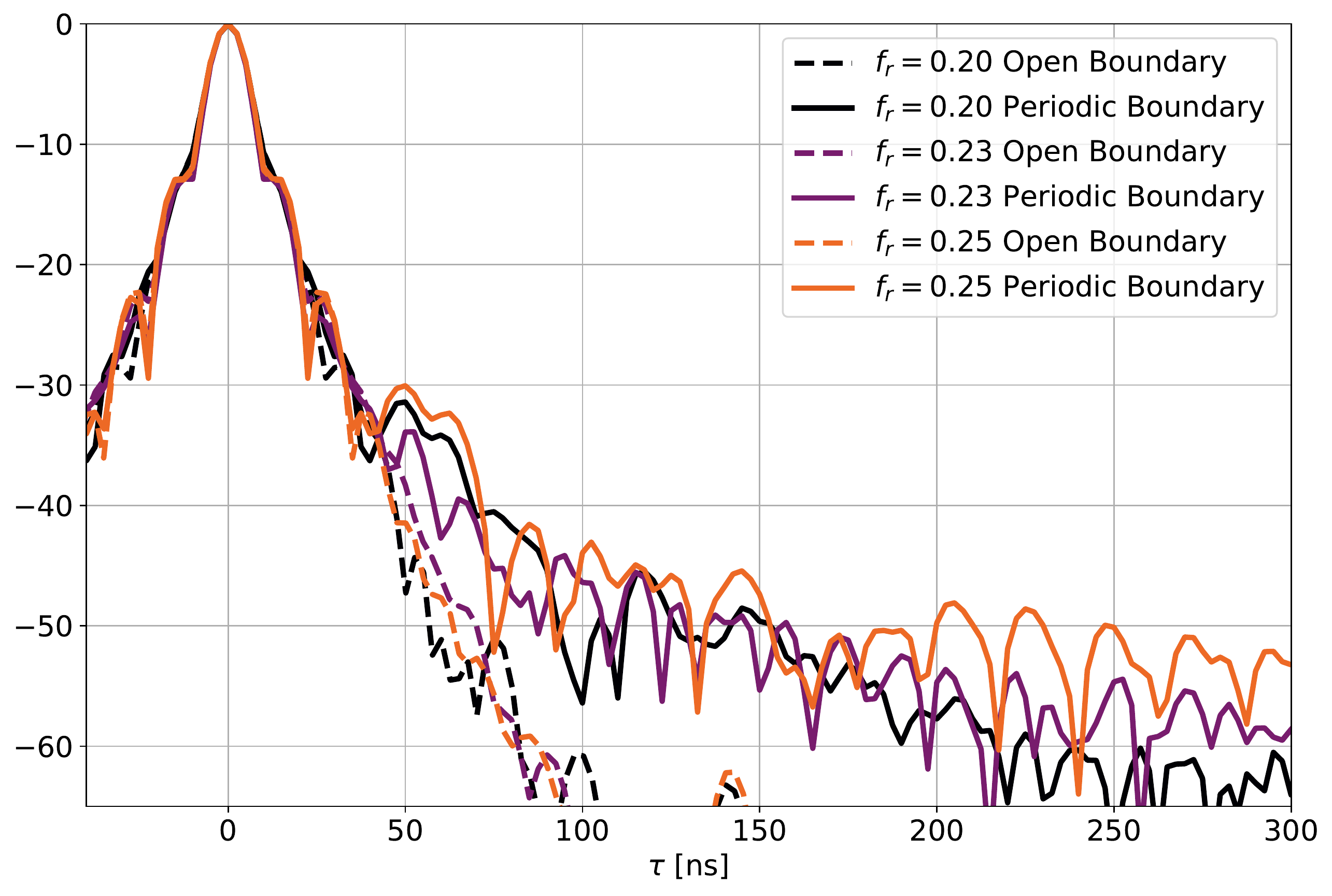}
    \caption{The delay kernel of the HIRAX dish towards a plane-wave at zenith with open boundary conditions (dashed lines) and periodic boundary conditions with two meters between dish edges (solid lines). This kernel determines the extent to which foregrounds are leaked from small line-of-sight Fourier modes where they exist intrinsically to large line-of-sight Fourier modes. In order to minimize foreground leakage and maximize our ability to recover cosmological fluctuations, this kernel should be kept as narrow as possible in delay. Different colors represent different values for the dish focal ratio for a fixed dish diameter of six meters. Reducing the focal ratio deepens the dish and lowers the feed below the rim. Comparing the open boundary curves, we see that reducing the focal ratio increases intra-dish reflections, raising the time-domain response by several dB. The curves with periodic boundaries include reflections of the sky signal off of nearby antennas. While a higher focal ratio is preferred for a single dish in isolation, we see that the inter-dish reflections dominate over intra-dish reflections, especially at higher delays. These inter-dish reflections are mitigated by lowering the focal ratio. Thus, when considering the HIRAX dish embedded in an array, we find that lowering the focal ratio has a net beneficial impact on the time-domain response. }
    \label{fig:f_ratio_delay}
\end{figure}

\begin{figure}[H]
\begin{center}
\includegraphics[width=0.45 \textwidth]{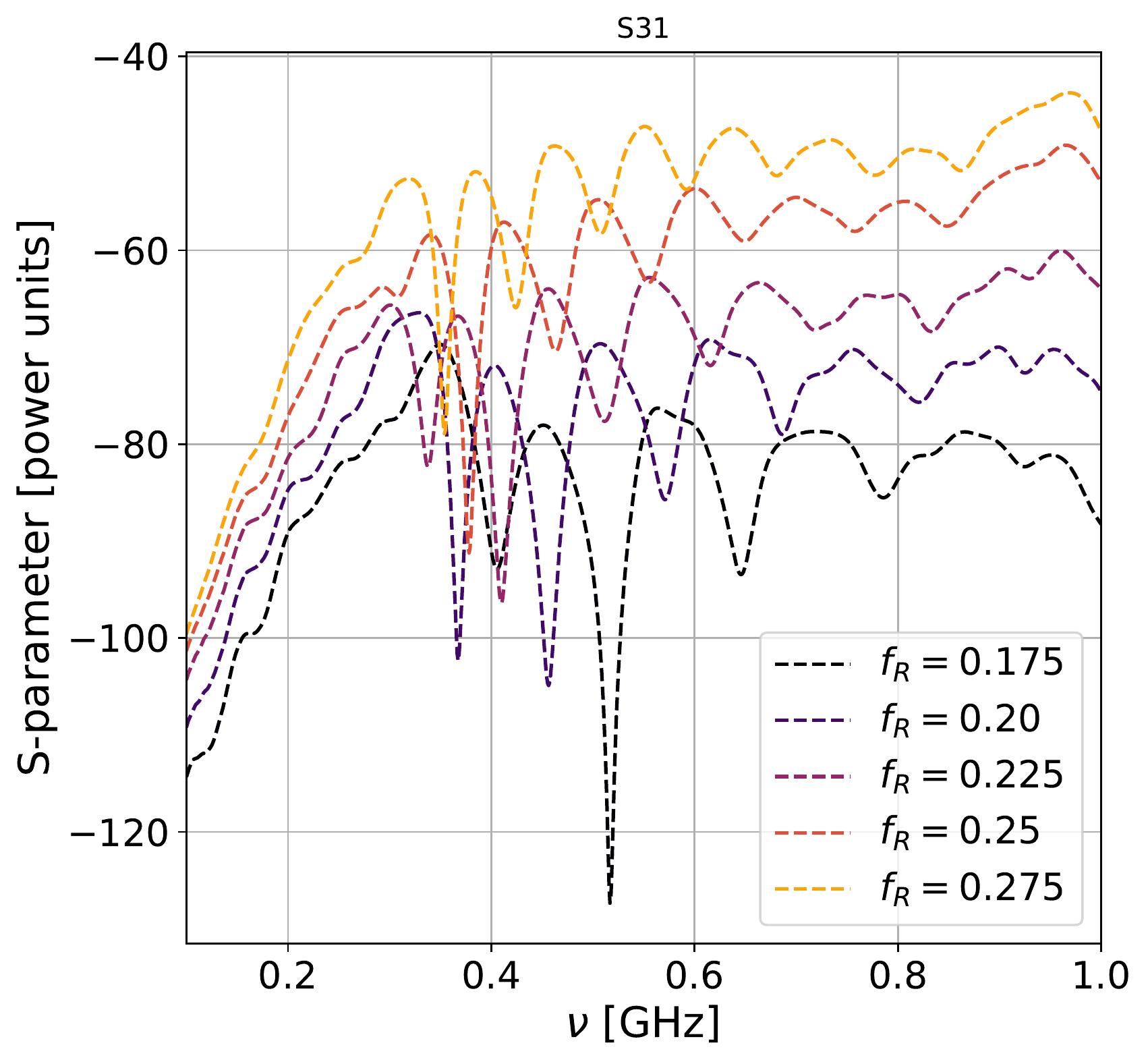}
\includegraphics[width=0.45\textwidth]{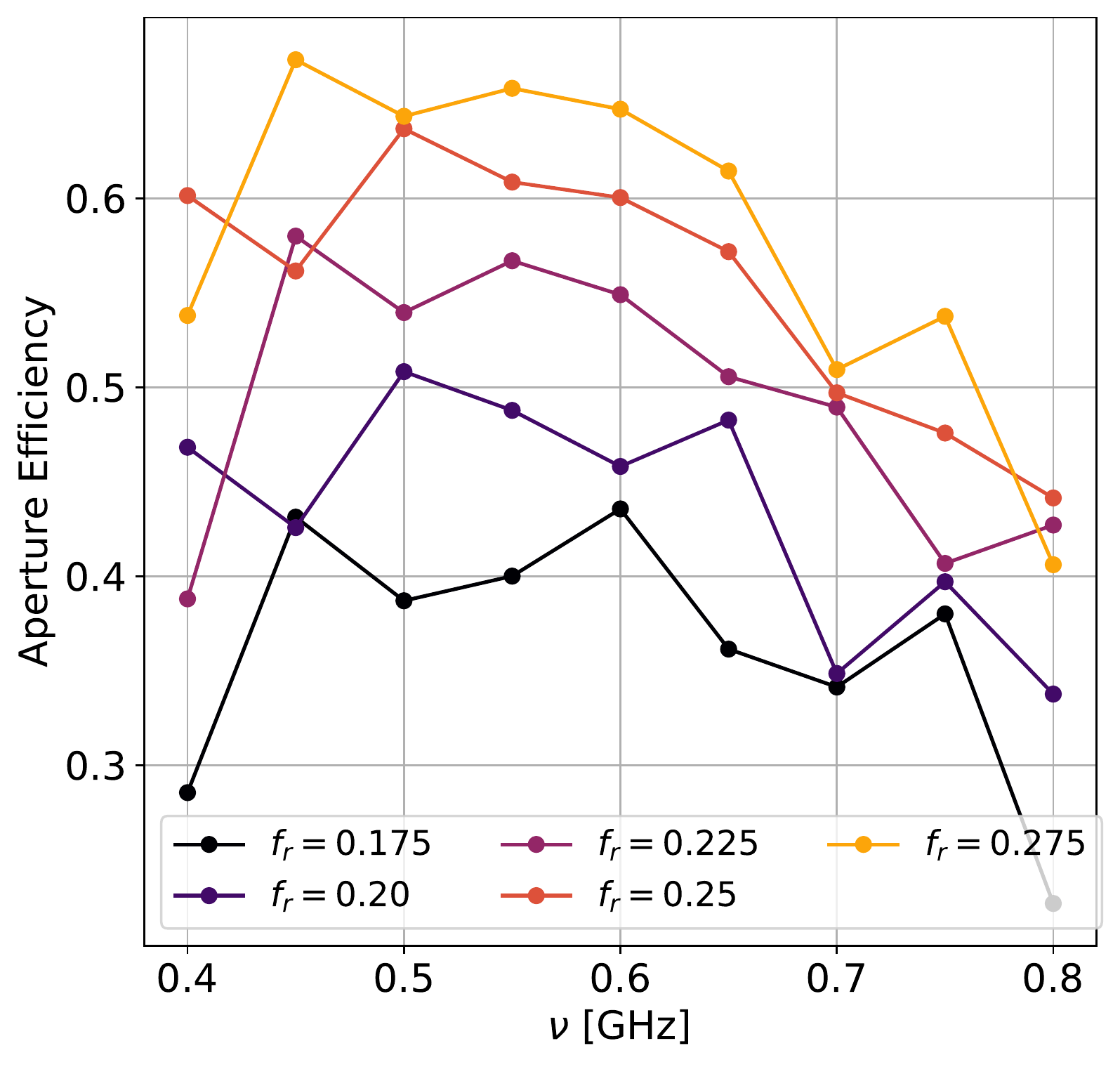}
\end{center}
\caption[]{Comparison of nearest-neighbor crosstalk ($S_{31}$, left), and aperture efficiencies (bottom), for a wide range of focal ratios. Decreasing focal ratio (deeper dishes) reduces crosstalk. Decreasing focal ratio also reduces the aperture efficiency (right), reducing overall system sensitivity. Geometries below f/D\aprx0.225 are also increasingly difficult to mechanically support, requiring significantly more backing structure due to the deep shape. The selected value of 0.23 reduces $S_{31}$ by $\aprx10$~dB relative to 0.25, while only reducing aperture efficiency by $\aprx5\%$.}
\label{fig:f_ratio_xtalk} 
\end{figure}

In Figure~\ref{fig:f_ratio_delay}, we compare the time-domain responses of antennas with different focal ratios with open and periodic boundary conditions. Comparing the open boundary curves, we see that decreasing the focal ratio worsens intra-dish reflections by several dB. However, comparing the curves from periodic boundary conditions, we see that inter-dish reflections dominate the time-response of a single element, especially at large delays, which are most sensitive to cosmology. From the perspective of reducing spectral structure, it makes sense for us to decrease the focal ratio for a fixed dish diameter as much as is allowed by mechanical constraints and cost. Reducing the focal ratio also reduces the potential for line-of-sight noise coupling by reducing $S_{31}$. We demonstrate this reduction in the right hand panel of Figure~\ref{fig:f_ratio_xtalk} where we plot the $S_{31}$ coupling parameter between two parallel polarized feeds of two adjacent antennas. 

However, reducing the focal ratio can also decrease the illumination of the dish and degrade sensitivity. In the bottom panel of Fig~\ref{fig:f_ratio_xtalk}, we see that the aperture efficiency of the dish is moderately affected by reducing the focal ratio. A focal ratio of 0.23 reduces $S_{31}$ by $\aprx10$~dB relative to 0.25, while only reducing aperture efficiency by $\aprx5\%$.
The selected focal ratio value of 0.23 was optimal for minimizing $S_{31}$ and the delay kernel amplitude at high delay ($\tau > 50$~ns), within the constraints of mechanical support and per-element cost. 

\subsection{Design Tolerance Simulations}
\label{subsec:spec_sim}

In addition to the design selection and optimization simulations, an extensive suite of simulations was performed exploring the results of inaccuracies in the manufacture or assembly of the instrument which degrade the redundancy of the array, in order to place specifications on the components and assembly. These simulations included, among others:

\begin{itemize}
    \item Varying feed position in six axes (translations and rotations)
    \item Varying dish diameter, holding focal ratio constant
    \item Varying dish focal ratio, holding diameter constant
    \item Perforations in dish conductive surface
    \item Reduced dish surface conductivity
\end{itemize}

The resulting beams from these simulations were in turn used to mock-observe simulated 21~cm skies, as described in Section \ref{sec:cosmo_sims}, in order to examine the effects of these mechanical errors on observations, and the recovered data products. In addition to non-uniform manufacturing, edge effects due to the finite extent of the array are also a source of non-redundancy. This category of effects is not explored here, but will be the subject of future work.

\subsubsection{Feed Position} For all simulations, the coordinate system origin is at the nominal feed position. The positive z-axis points from the feed to the dish vertex, the x-axis is horizontal when the dish is pointed at the horizon, and the y-axis is vertical. For the feed position simulations an array of simulations were performed for each axis, shifting the feed along that axis. The positions were: from 0~mm to 10~mm from the nominal focal point in 1~mm steps along the x and y-axes, and from -5~mm to 5~mm in 1~mm steps along the z-axis. For the three rotational axes, the feed was rotated from $0^\circ$ to $5^\circ$ in $0.5^\circ$ increments, around each axis.

The primary effect of translation along the z-axis is to throw the instrument out of focus, which reduces gain, and alters beam sidelobe structure. Translations in the other two dimensions primarily result in beam pointing errors, and asymmetric sidelobes. Rotation around the z-axis does not change the shape of the beam, but due to the non-azimuthally symmetric sidelobe does result in relative changes in sidelobe amplitudes between pairs of dishes with different rotations. Rotations around the x and y-axis primarily change the pointing of the beam, and creates asymmetric sidelobes. The resulting design tolerance was $\pm1$~mm in all linear translation directions, $\pm1.5$~arcmin for the azimutal rotation, and $\pm2.5$~armin for the other rotations. The summary of design specifications can be seen in Table \ref{tab:specs}.

\subsubsection{Dish Diameter} The dish diameter was varied from 599~cm to 601~cm in 1~mm steps, while holding the focal ratio constant (so as not to conflate the results with those of the focal ratio simulations). Changing the dish diameter throws the instrument out of focus, and changes the beam width. We determined the accuracy, averaging over they array, must be within $\pm3$~mm, and the individual dish precision within $\pm1$~mm.

\subsubsection{Focal Ratio} The focal ratio was varied from f/D = 0.245 to 0.255 in steps of 0.001, while holding the dish diameter fixed at 6~m. Changing the focal ratio changes the aperture efficiency, and can alter the sidelobe levels. These simulations were distinct from those exploring the focal ratio in Section \ref{sec:design_sims}, as they were intended to explore fine variations around a nominal value, to constrain manufacturing tolerances, not to explore large changes in the nominal value. These simulations contributed to the dish shape accuracy specification in Table \ref{tab:specs} ($\pm3$~mm maximum deviation from the ideal paraboloid curve).

\subsubsection{Dish Perforations} Radio dishes are often constructed of metal mesh instead of solid metal panels, for purposes of lightweighting, lowering wind cross section, and cost reduction. As long as the gaps in the mesh are substantially smaller than the wavelength, there is no reduction in the performance, but as the gap size approaches the wavelength, the dish becomes increasingly transparent. A common criteria given for maximum gap sizes is $\lambda/10$. We explored the effects of hole size on the instrument beams in order to set specifications on what types of mesh were permissible and to regulate production errors resulting in accidental perforations or tears in the dish surface. As a first-pass method of simulating this effect, the dish was perforated with a regular array of circular holes, arranged in 16 equally spaced azimuthal angles ($22.5^\circ$ separation), and 6 radial distances from the center of the dish with 25~cm spacing, for a total of 96 perforations in the dish. The hole diameter was varied from 0.5~cm to 5.0~cm, in 5~mm steps.

This method was used because a realistically large number of holes in the dish would vastly increase the complexity of the meshing, and therefore the simulation run times. The holes were restricted to a central 1.5~m radius of the dish to again simplify the model: at this diameter holes that are circular in cross section and projected from a plane normal to the dish vertex are not significantly distorted by the curvature of the dish. If holes were projected out to the full 3~m radius of the dish, they would be extremely elongated in the azimuthal direction, and instead of exploring one hole radius, we would be exploring a radially and azimuthally varying range of values. Altering the projection angle of the holes with the radius would allow the full surface area to be perforated, and will be explored in future analysis analysis. 

Perforations create additional structure within the beam, which increases with increasing hole size, and at higher frequencies. At larger hole sizes it decreases gain, and increases power from the ground (though this effect was not modeled in our simulations). The resulting specification on perforation size from this analysis (5~mm maximum gap dimension, see Table \ref{tab:specs}) is significantly stronger than the $\lambda/10$ criteria (as large as 7.5~cm at 400~MHz). In future this set of simulations will be expanded to a more realistic model using a pseudo-randomly placed array of perforations over a wider area of the dish.

\subsubsection{Conductivity} Lastly, we simulated different conductivity values of the dish's reflective surface to determine any impact on the beams (if for example, the metal mesh used in the surface was excessively fine, and the resistance per strand increased); or if a non-uniformity in the conductivity would have an effect on the beams (as if the mesh density varied, or if different metals were used in different parts of the dish). Two sets of simulations were run to explore these possibilities, one in which the conductivity of the entire dish was varied, and one in which the dish was modeled as two halves with different conductivities. The latter represents a limiting case, and the beam effects calculated from a dish surface with maximally inhomogeneous conductivity can be interpreted as an upper limit. In the first set of simulations, the conductivity was varied from $6 \times 10^7$~S/m (approximately the conductivity of copper) to $1 \times 10^6$~S/m (roughly the conductivity of stainless steel, and the lowest conductivity of materials one might reasonably use) in steps of $5 \times 10^6$~S/m. In the second set of simulations, one half of the dish was held at $6 \times 10^7$~S/m, while the other was stepped down from $6 \times 10^7$~S/m in increments of $5 \times 10^6$~S/m to $1 \times 10^6$~S/m. This series of simulations showed essentially no changes in the beam over the range of explored values. This is in agreement with experimental measurements of the radiation efficiency of dipole and meander antennas as a function of conductivity, which show no change in efficiency until the conductivity is below $1 \times 10^6$~S/m \cite{shahpari15}. The model of conductivity variations employed in these simulations is a simplified model of the large possible parameter space of variations in the dish surface conductivity, and will be expanded upon in future simulations.

\begin{table*}
\begin{center}
{
\caption{HIRAX Telescope Element Specifications}
\begin{tabular}{l|c|l}
\hline \hline
 Element & Specification & Notes \\
\hline
 Axial symmetry of & $\pm$1~mm &  \\
 receiver support & & \\
\hline
 Receiver support & $< 0.5$ dB & \\
 RF attenuation & & \\
\hline
 Deviation of power & $\pm$2~mm & \\ 
 cabling from boresight & & \\
\hline
 Rigidity of & \ $\pm0.5$~mm & In x,y, and z dimensions \\
 receiver support & & \\
\hline
 Positioning of receiver & $\pm0.5$~mm & In x,y, and z dimensions \\
 relative to focal point & & \\
\hline
 Orientation of receiver & $\pm2.5$~arcmin  & polar angle \\ 
 relative to boresight & $\pm1.5$~arcmin &  azimuthal angle\\
\hline
 Dish diameter & $\pm3$~mm & Accuracy \\
 & $\pm1$~mm & Precision \\
\hline
 Dish shape accuracy & $\pm3$~mm & Deviation from ideal paraboloid \\
\hline
 Dish electrical connectivity & $< 5$~mm & Maximum dimension of gaps \\
\hline
 Dish surface conductivity & $> 1 \times 10^6$ S/m & \\
\hline
\label{tab:specs}
\end{tabular}}
\begin{tablenotes}
Instrument specifications determined by EM simulations. This is a subset of the total system specifications.
\end{tablenotes}
\end{center}
\end{table*}

\section{Photogrammetry Measurements}

The specifications outlined above are significantly more stringent than those typically required for radio telescopes at these wavelengths, due to the increased accuracy necessary to control array redundancy. Achieving these specifications within cost is a significant engineering challenge. Photogrammetry measurements of our first prototype dishes were performed to ascertain whether the specifications had been meet, and to inform the design and production of future dishes. The results show that already in the first prototype we are close to achieving the specified goals, though improvements still need to be made.

\begin{figure}[H]
\begin{center}
\includegraphics[width=0.8 \textwidth]{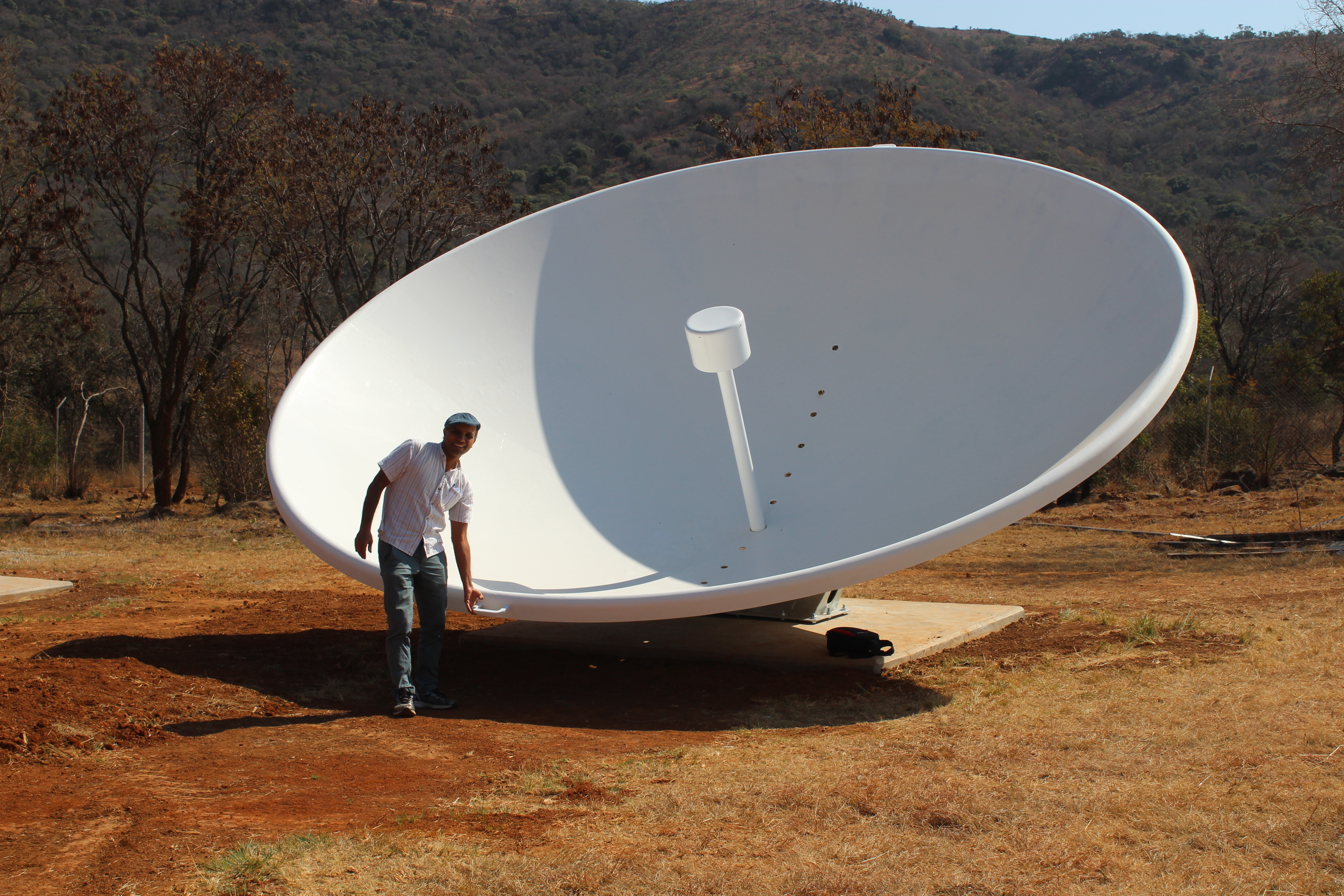}
\end{center}
\caption[]{A 6~m diameter f/D = 0.25 prototype dish at the Hartebeesthoek Radio Astronomy Observatory in South Africa. The dish is aluminum embedded in fiberglass composite, and the receiver is supported from the dish vertex by a 10~cm diameter fiberglass column. HIRAX will operate in drift scan, so the mount has only an elevation axis, which allows for pointings between $\pm30^\circ$ degrees off zenith. PI for scale.}
\label{fig:dish} 
\end{figure}

The depth of the dishes is designed to reduce crosstalk between the elements in the close-packed array, but this is an uncommon design for radio telescopes (though similarly deep dishes have been used for other close-packed radio arrays such as HERA\cite{deboer17}, CHIME\cite{bandura14}, and CHORD\cite{vanderlinde19}), and requires greater support than more typical shallow profile dishes. 
Figure \ref{fig:dish} shows a prototype dish located at the Hartebeesthoek Radio Astronomy Observatory (HartRAO). The pictured prototype dish was manufactured by MMS Technology Ltd. in South Africa (mmstechnology.co.za). A second metal dish prototype was constructed jointly by NVJ and Rebcon Engineering, but is not discussed here.

\begin{figure}[H]
\begin{center}
\includegraphics[width=0.45 \textwidth]{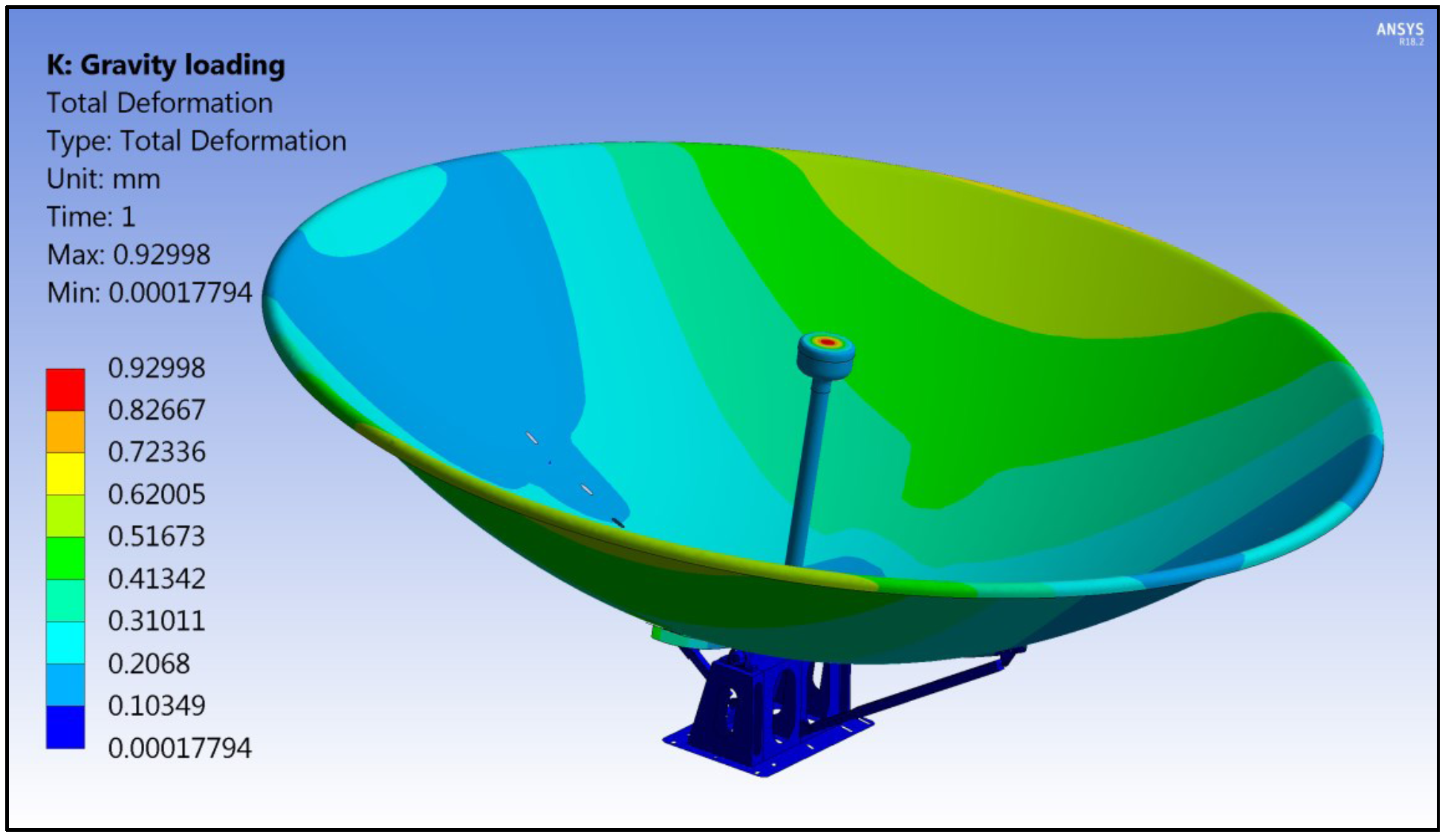}
\includegraphics[width=0.45 \textwidth]{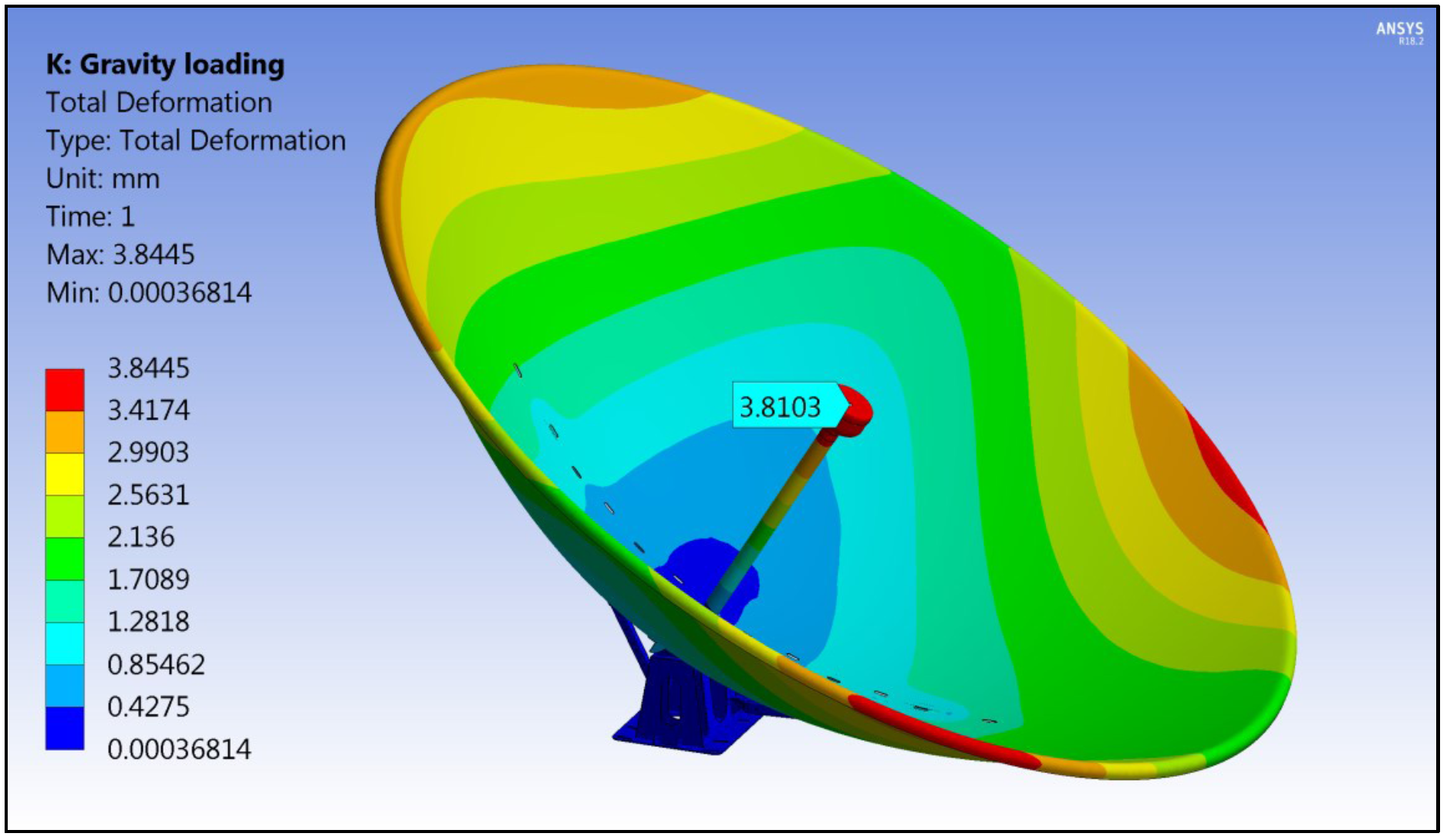}
\includegraphics[width=0.45 \textwidth]{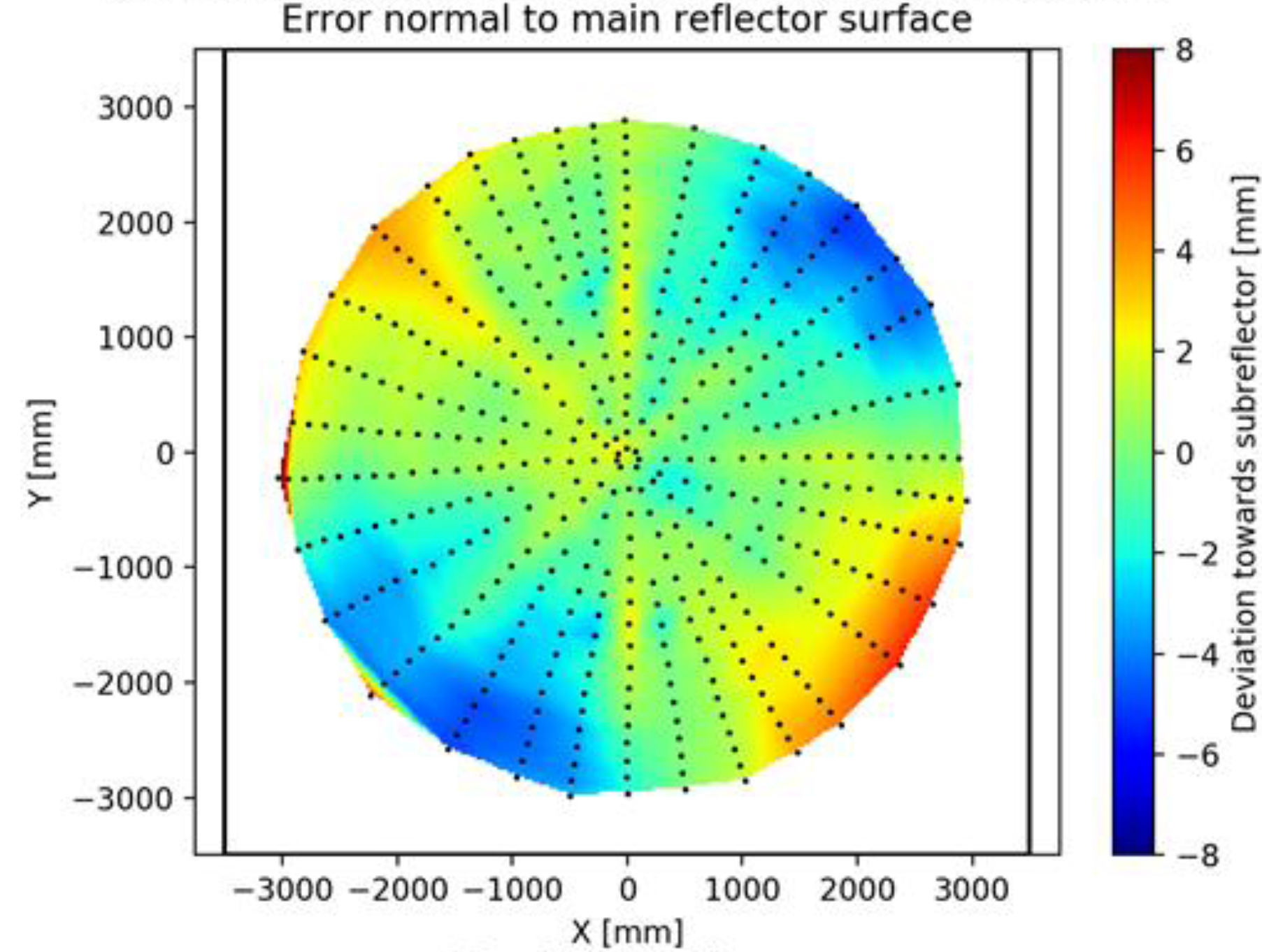}
\includegraphics[width=0.45 \textwidth]{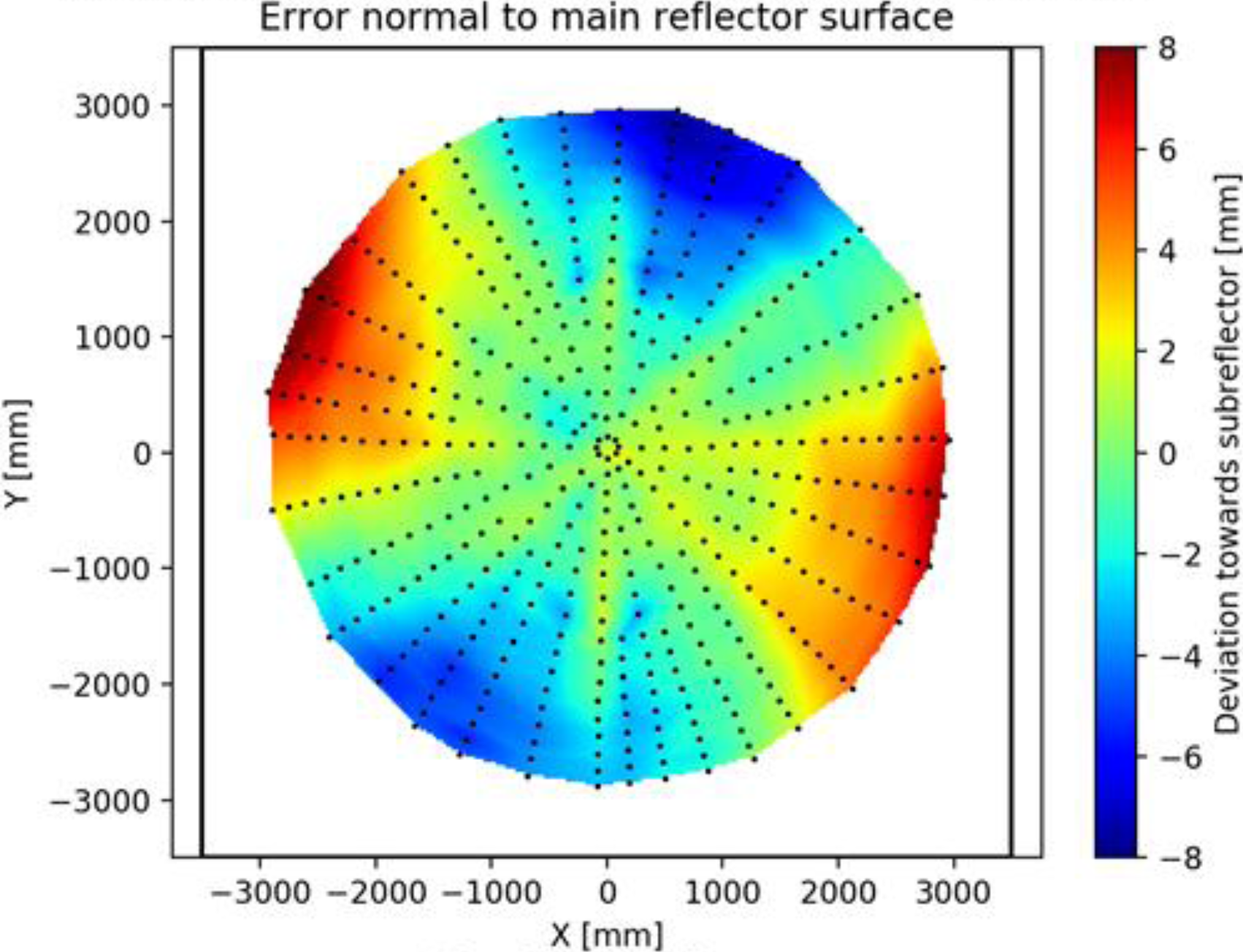}
\end{center}
\caption[]{HIRAX prototype dish photogrammetry and FEA simulations. The top row shows FEA simulations with predicted total deformations, while the bottom row shows the measured deviations from the theoretical paraboloid surface, measured normal to the surface. The left column shows zenith pointing, the right column shows $30^\circ$ elevation below zenith, the lowest pointing for the HIRAX survey. 
The measured deformations, while close to the design specifications, are several times larger than those predicted by the simulations. 
Dish FEA simulations were performed by MMS Technology (Pty) Ltd., and figures are courtesy of Heinrich Bauermeister. Photogrammetry was performed by the South African Radio Astronomy Observatory (SARAO), and figures are courtesy of Mattieu de Villiers.
}
\label{fig:dish_photogrammetry} 
\end{figure}

Photogrammetry measurements were performed on the prototype dish in Figure \ref{fig:dish} to assess the accuracy of the dish surface and to compare gravitational deflections to those predicted by mechanical finite element analysis (FEA) simulations. Figure \ref{fig:dish_photogrammetry} shows FEA simulations\footnote{Dish FEA simulations performed by MMS Technology (Pty) Ltd. Figure courtesy of Heinrich Bauermeister.} of the f/D = 0.25 dish, and corresponding photogrammetry results\footnote{Photogrammetry performed by the South African Radio Astronomy Observatory (SARAO). Figure courtesy of Mattieu de Villiers.} for zenith angles of $0^\circ$ and $30^\circ$. 

The photogrammetry data points are fit to a rotationally symmetric paraboloid, and the plots show the data residuals with respect to this fit.  The dominant deformation is the quadrupolar ``potato chip'' mode, which arises from the cross-shaped dish backing structure and increases in amplitude with zenith angle.
The FEA simulations confirm the quadrupolar pattern of the distortions under gravitational load; however, the expected amplitudes are several times smaller than those observed in measurements. Deformations at zenith pointing are predicted to be less than 1~mm, but are measured to have a maximum amplitude of 5~mm. At $30^\circ$ zenith angle the maximum amplitude increases to 4~mm in the model and 8~mm in the measured data.

The deformations in both the zenith and $30^\circ$ pointing are also most significant towards the edge of the dish. This may be an indication that a more substantial backing structure is necessary to support the large diameter of the dish.

\begin{figure}[H]
\begin{center}
\includegraphics[width=0.3\textwidth]{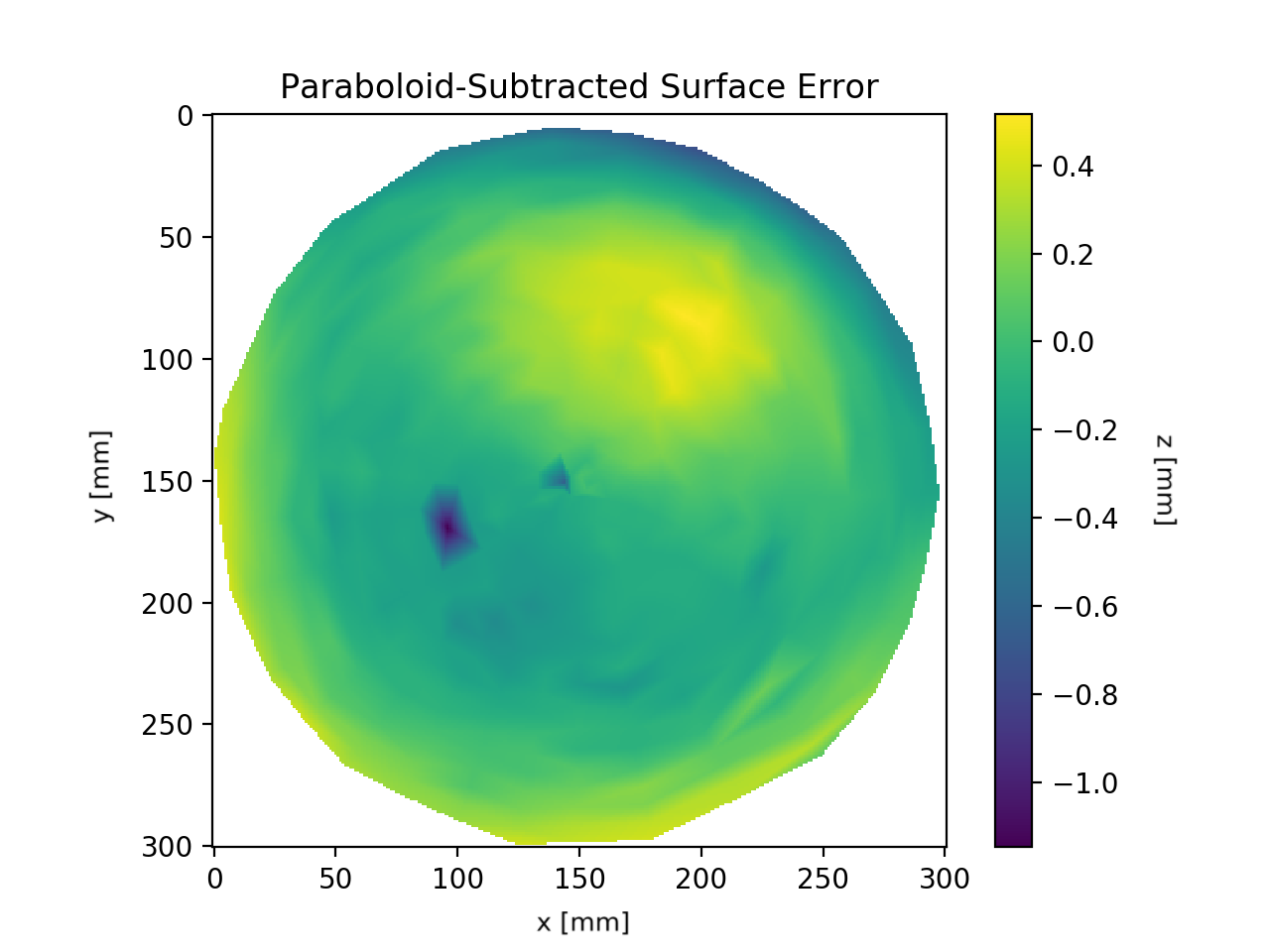}
\includegraphics[width=0.3\textwidth]{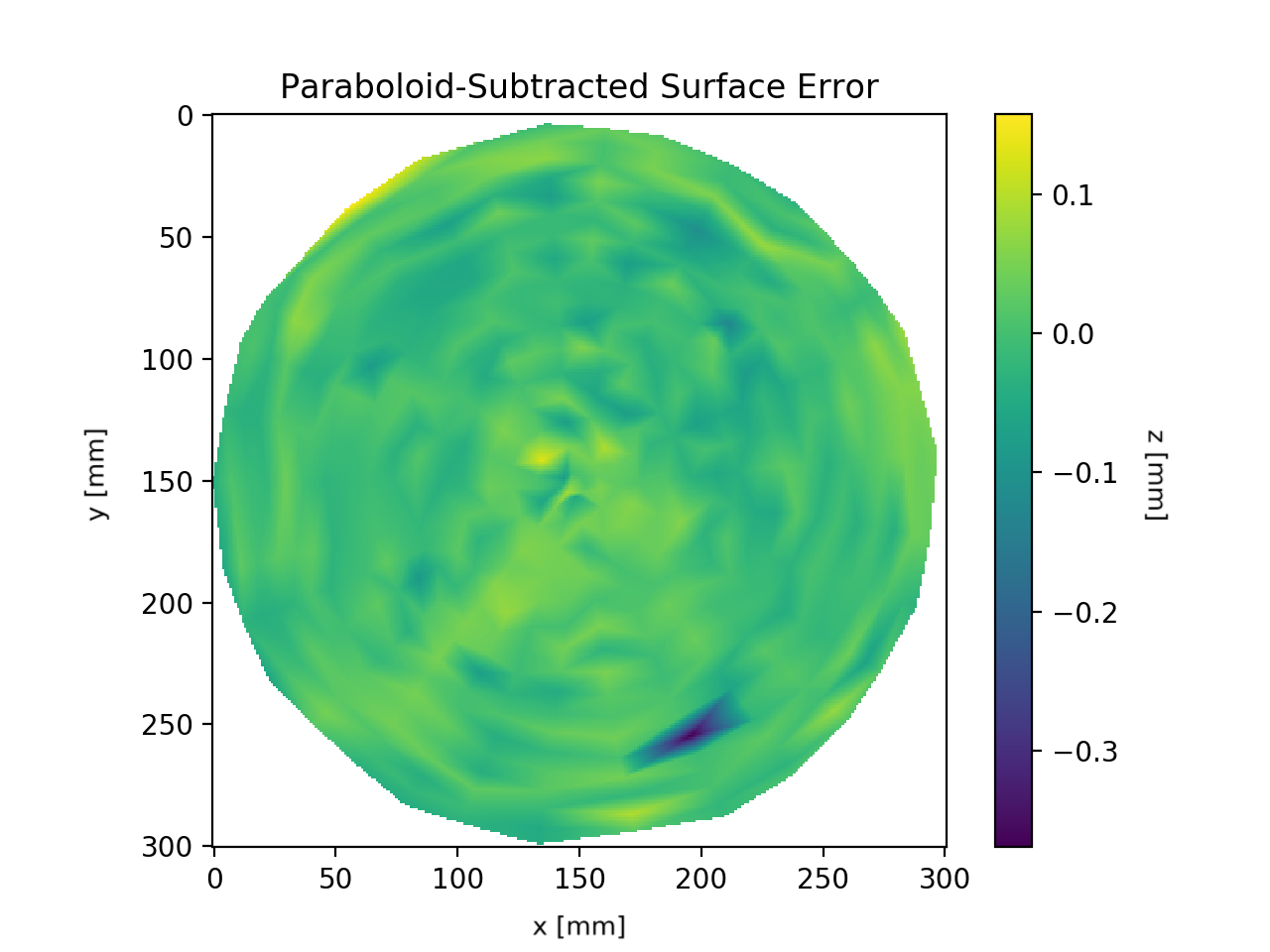}
\includegraphics[width=0.3\textwidth]{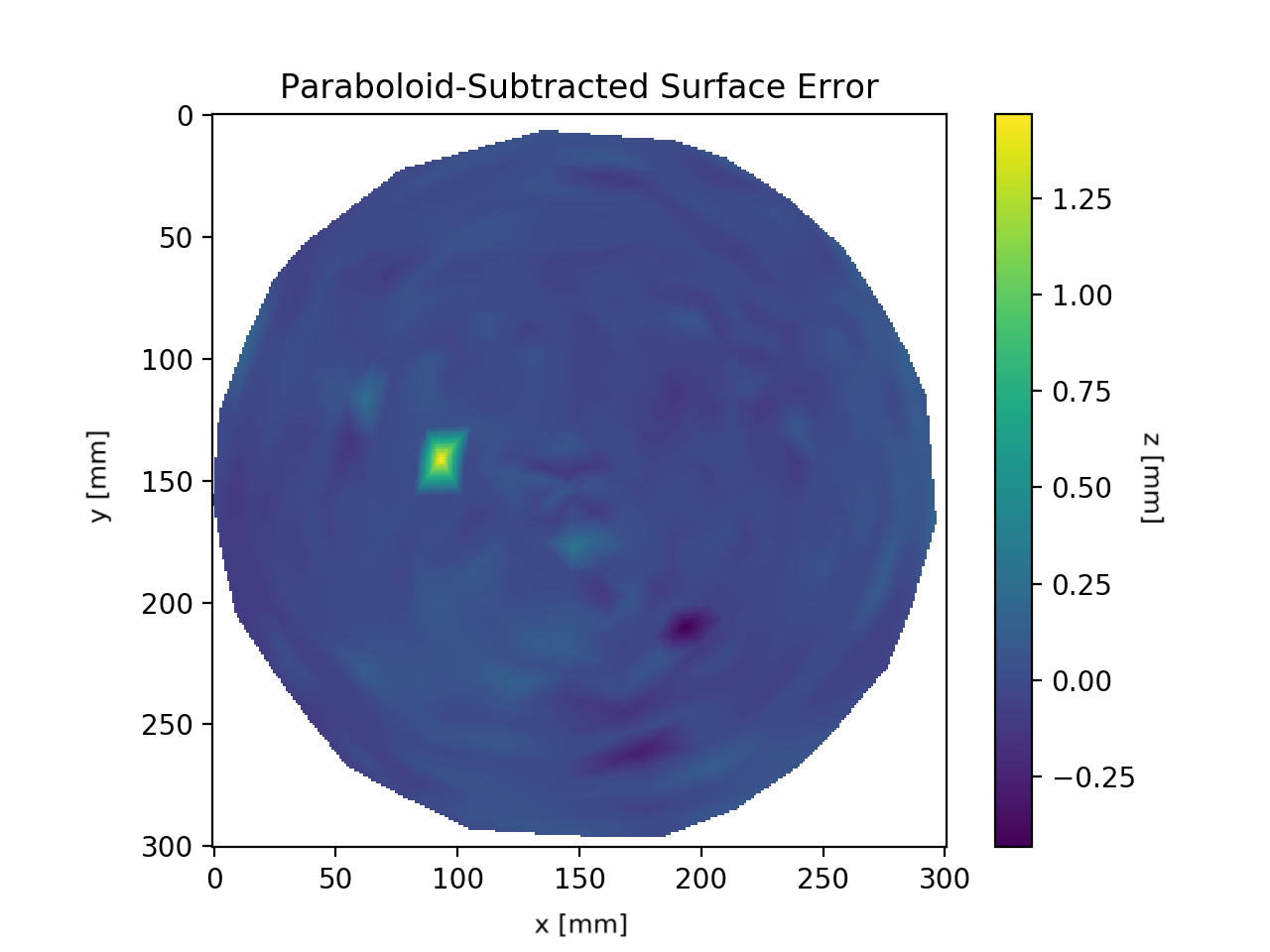}
\includegraphics[width=0.3\textwidth]{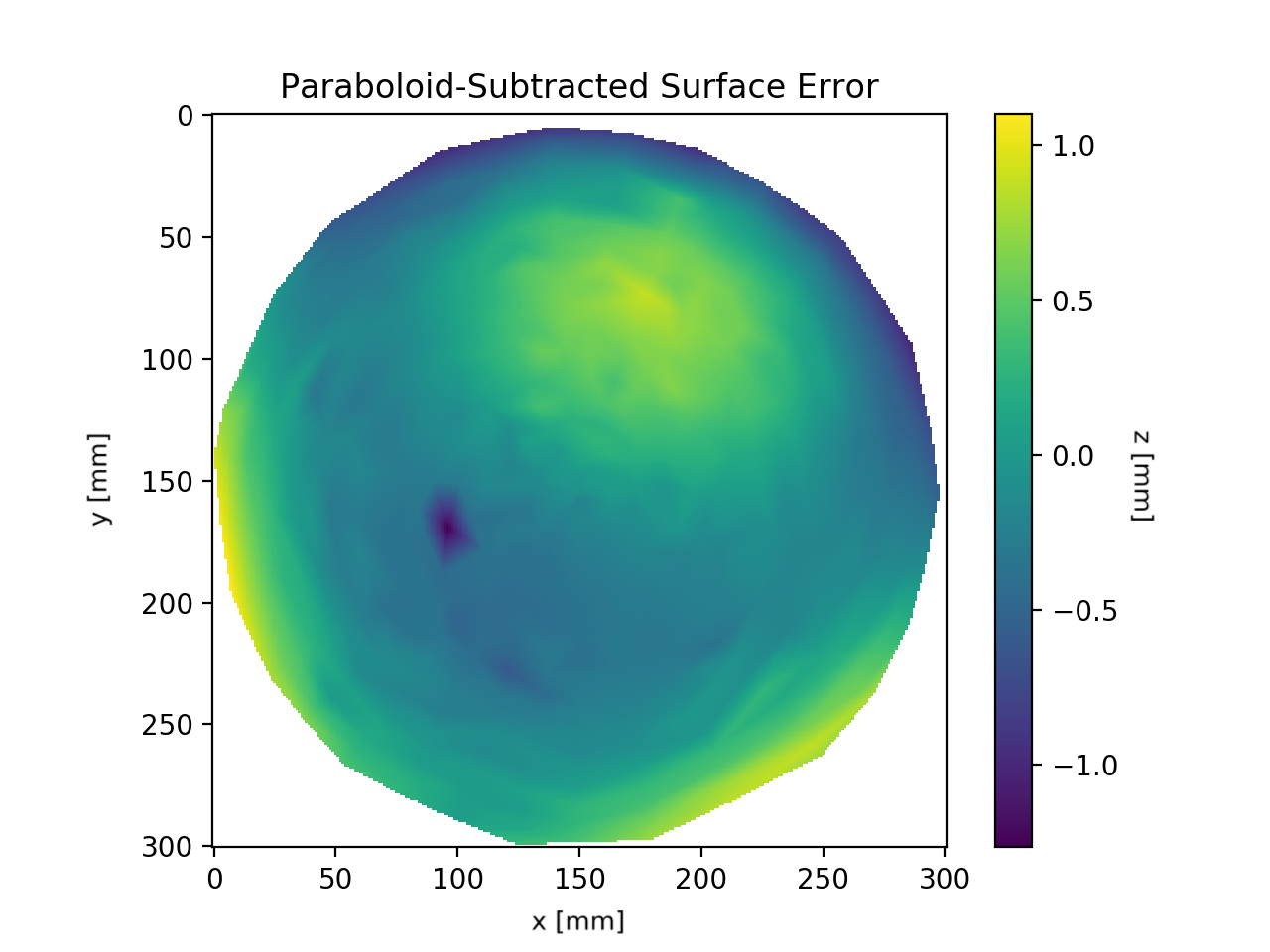}
\includegraphics[width=0.3\textwidth]{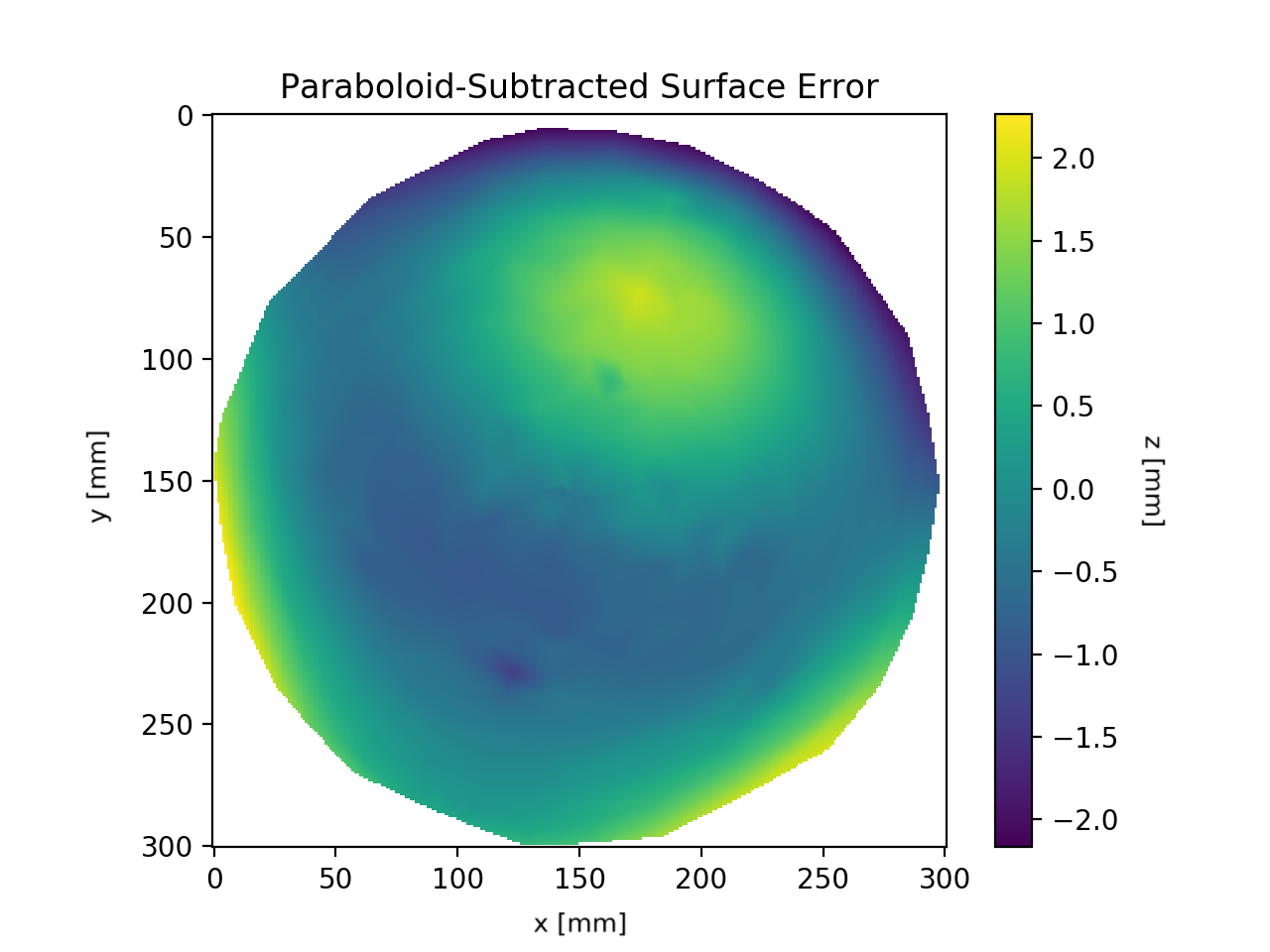}
\includegraphics[width=0.3\textwidth]{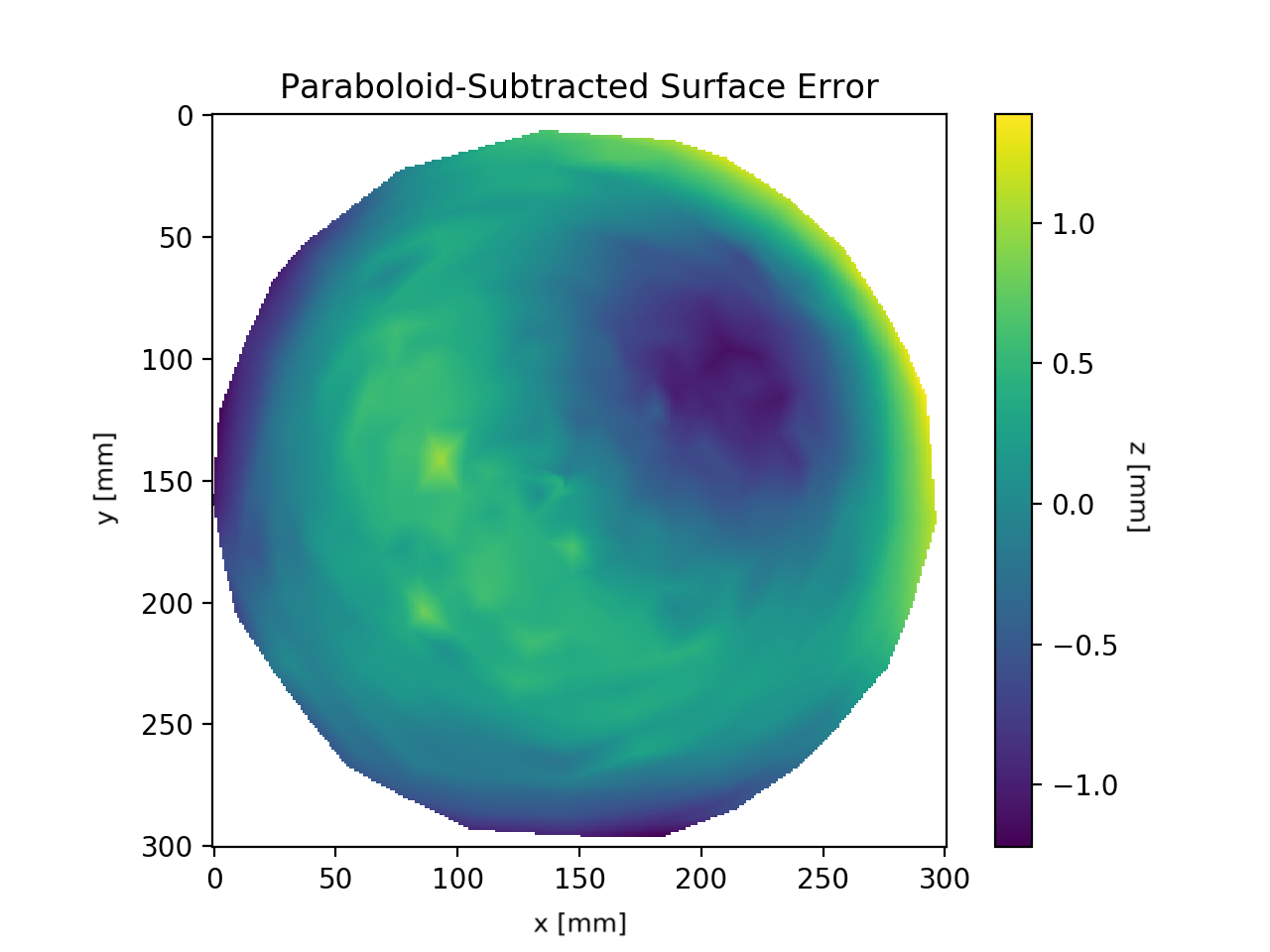}
\end{center}
\caption[]{Residual displacements after subtracting best-fit paraboloid and quadrupole. Color scales are in millimeters. Top row (left to right): $0^\circ$ (zenith), $15^\circ$, and $30^\circ$. Deviations from the model are below 1~mm, showing that the surface is well described by a paraboloid plus a quadrupolar distortion arising from the backing structure. Bottom row, differences between residuals at angle pairs, showing the systematic changes in the dish surface as the dish is tilted (left to right): $0^\circ$ and $15^\circ$, $0^\circ$ and $30^\circ$, and $15^\circ$ and $30^\circ$. 
}
\label{fig:model_subtracted_dish}
\end{figure}

Figure \ref{fig:model_subtracted_dish} shows the dish surface displacements after the best-fit paraboloid and quadrupole are subtracted at three different angles: $0^\circ$ (zenith), $15^\circ$, and $30^\circ$. The residual displacements are less than 1~mm, showing the dish surface shape is well described by the combination of the ideal paraboloid and a quadrupolar mode. Figure \ref{fig:model_subtracted_dish} also shows the differences between pairs of angles, which shows the systematic changes in the dish surface as the elevation angle is adjusted. These differences are larger, with maximum amplitudes of between one and two millimeters. This is further evidence that the dish needs addition support structure, to reduce the variations in dish shape as it is tilted.

This data informs our future design and production methods; while close to acceptable limits, the dish surface needs to be more precise, and the backing structure may need to be improved as well. The dish at zenith has a measured RMS surface deviation of 2.8~mm, and the requirement is $<1$~mm RMS deviation from the ideal paraboloid surface. We are confident that the mechanical improvements required to achieve the design specifications are feasible. The derivation of the specifications is discussed below.  

\section{Implications for 21~cm Cosmology}
\label{sec:cosmo_sims}

In order to examine the effects of beam shape systematics related to the mechanical specification of the dishes, we have propagated visibility perturbations derived from the CST simulations described above through to 21~cm power spectrum errors. The method for this is outlined briefly here but will be more comprehensively described in future work.

Firstly, we construct a summarised representation of the deviations observed in the beam shapes in the CST parameter sweep simulations outlined in Section~\ref{subsec:spec_sim}. This is done by applying a principle component analysis (PCA) to the observed radial profile deviations of the beam directivities exported from the CST simulations (currently only deviations in the co-pol. beams are considered). From this analysis, we select the 2 modes (functions of frequency and radial angular coordinate) that encapsulate the most variance in the beam radial profile for a given parameter in the dish-feed system that has been swept. The appropriately weighted linear combination of these modes then represents an approximation of the linear-order perturbation of the beam shape, which can be expressed as a derivative with respect to the physical parameter under consideration. With this linear approximation of the effect of varying a given physical parameter on the beam shape, we can efficiently generate realisations of perturbed beams motivated from what is observed in the CST simulation parameter sweeps.

\begin{figure}[H]
\begin{center}
\includegraphics[width=0.45\textwidth]{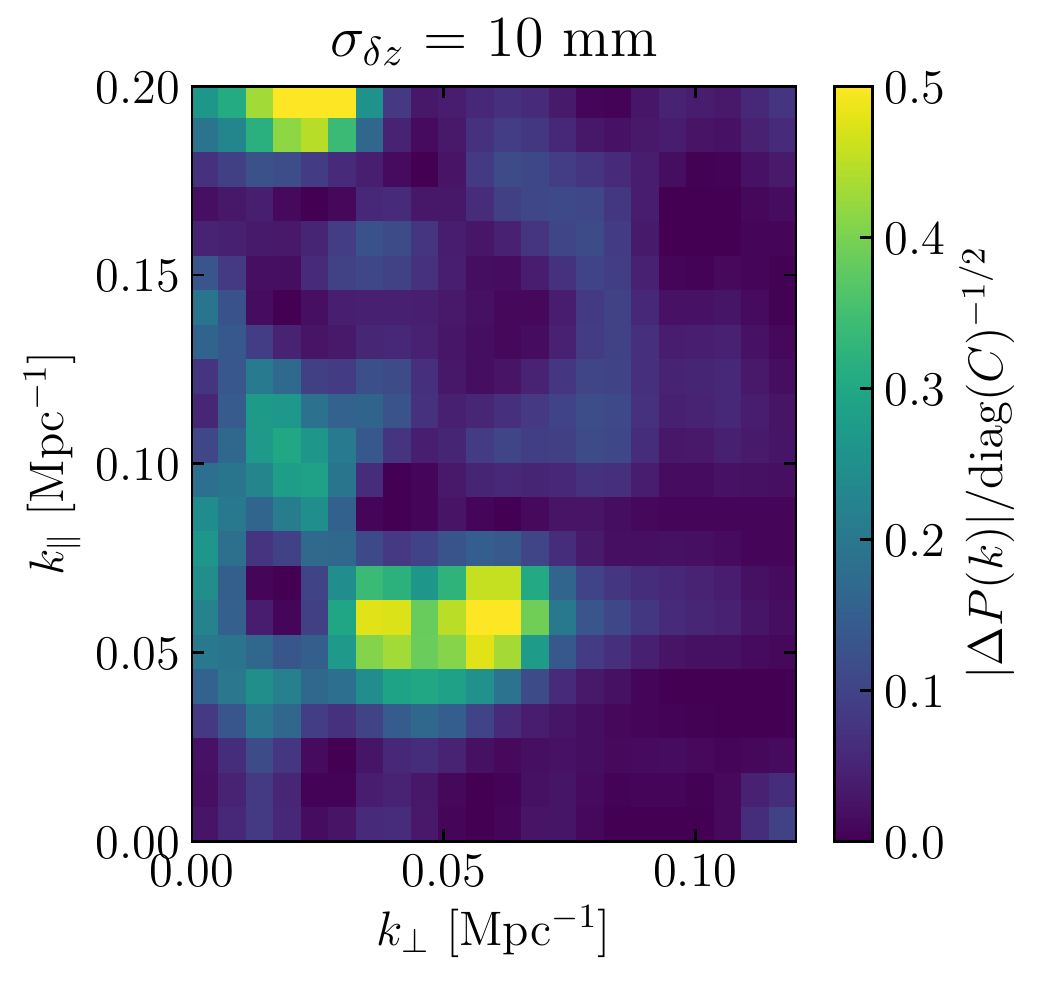}
\end{center}
\caption[]{Deviations in the recovered 21~cm power spectrum due to a single realization of systematic beam perturbations due to Gaussian distributed positional offsets (with standard deviation=10~mm) of the feed along the focal axis, as calibrated by CST simulations. The results shown here are scaled by the estimated statistical noise on the power spectrum bins such that values of magnitudes greater than one would indicate systematic contributions larger than the estimated statistical noise level. The simulations from which these results were generated consider only the frequency range of 600-650~MHz and utilises only a subset of the baselines available to HIRAX, but assumes the same level of redundancy as in the 1024 element array. Upcoming publications will expand upon these results.\label{fig:hirax_ps_sys} }
\end{figure}

To evaluate a potential systematic's effect on 21~cm cosmology, we generate random realisations of the perturbed beams based on a distribution of physical parameters across the dishes of the array. Figure~\ref{fig:hirax_ps_sys} shows the effect of such perturbations in the feed position along the focal axis relative to the dish, $\delta z$, on the recovered 21~cm power spectrum. Here we have assumed the distribution of $\delta z$ across dishes is that of a Gaussian with standard deviation of 10~mm. Using the perturbed beams we first generate visibility-space perturbations in simulated data assuming a nominal sky model including galactic foregrounds and cosmological 21~cm signal, and a nominal instrument and survey. These perturbed visibilities are then used to construct a foreground filtered power spectrum estimate using a modified version of the $m$-mode formalism \cite{shaw14, shaw15}. In Figure~\ref{fig:hirax_ps_sys} a plot of the deviations of the recovered power spectrum with respect to the input 21~cm power spectrum due to these systematic perturbations is shown.

Currently this effort has only considered small scale simulations of the array (limited to core baselines and subsets of the full frequency range), and radially symmetric perturbations in the co-pol. beam shapes. Future work will extend this to full scale simulations using a less summarised prescription for the beam deviations as well as incorporating full polarization information.

These results have been used to inform the telescope specifications listed in Table~\ref{tab:specs}. For this purpose we first determined our tolerance on systematic contributions to the required power spectrum sensitivity from high-level Fisher matrix forecasts on Dark Energy parameter constraints. This tolerance was set such that the impact of the systematic effects under evaluation would reduce the Fisher-forecasted Dark Energy Figure of Merit (FoM) by no more than 20\% of the fiducial forecasted level that assumes a power spectrum noise comprised of purely statistical noise with a nominal residual foreground contribution. Table~\ref{tab:specs} represent a mechanically achievable set of specifications that is consistent with this tolerance as informed by the simulations described here. A more quantitative description of this process will be presented in future work.

\section{Summary and Future Work}

The simulations and verification measurements performed for this work have significantly impacted the design of the HIRAX array. The cabling and receiver support structure were completely re-designed to improve RF performance and mechanical stability. Routing the cabling precisely along the telescope axis of symmetry reduces sidelobe amplitudes by 7~dB from the previous prototype design, and eliminates asymmetries in the sidelobe structure. Selecting a large (35~cm) diameter receiver support column over a small (10~cm) support column increases the main beam amplitude by 0.5~dB. The focal ratio was adjusted, resulting in a significant decrease in crosstalk. Decreasing the focal ratio from 0.25 to 0.23 reduces inter-dish reflections by up to 5~dB, as measured by the delay kernel, and line-of-sight noise coupling ($S_{31}$) by up to 10~dB, while only reducing aperture efficiency by $5\%$, from approximately $60\%$ to $55\%$. We have also conducted extensive range measurements of the HIRAX feed to verify the accuracy of our simulations, and are preparing for field measurements of the full 6~m dish elements with a drone based calibration platform. 

We have traced the effects of the modeled system non-redundancies to recovered cosmological parameters, and have derived specifications for the manufacture and assembly of the dish from those cosmological simulations, as summarized in Table \ref{tab:specs}. This is of particular importance for the 21~cm cosmology field: this is the first time the redundancy requirements of a 21~cm array have been explicitly derived from the telescope mechanical and electromagnetic performance to the cosmology simulations pipeline.

Future simulations will continue to examine the effects of intra-dish and inter-dish reflections on array performance, will expand the simulation models of dish perforations and conductivity beyond the first-pass models used here, and examine the effects of large scale (quadrupolar) and small scale (RMS) deformations of the dish surface on the instrument beams and survey cosmological constraints. Future publications will expand on the details of the HIRAX electromagnetic simulations, beam verification measurements, and cosmological simulations. There is currently a tender out for production of the dishes for the HIRAX 256-element array, based in part on the specifications derived here, and construction of the array will begin once the tender is complete, in 2021. 

\acknowledgments 
 
This work was supported by the National Science Foundation (NSF) under Grant No. 1751763. AEW acknowledges support from the Berkeley Center of Cosmological Physics. ERK is supported by a NASA Space Technology Research Fellowship (NSTRF). KM acknowledges support from the National Research Foundation (NRF) of South Africa. The HIRAX radio array is funded by the NRF. DC acknowledges the financial assistance of the South African Radio Astronomy Observatory (SARAO, www.sarao.ac.za). Results presented here were produced with the support of the computing and technical resources of the Yale Center for Research Computing and Wright Laboratory, and with the support of personnel and facilities at NCSU, Caltech, and NASA JPL.

\bibliography{saliwanchik} 

\begin{thebibliography}{10}

\bibitem{wein13}
{Weinberg}, D.~H., {Mortonson}, M.~J., {Eisenstein}, D.~J., {Hirata}, C.,
  {Riess}, A.~G., and {Rozo}, E., ``{Observational probes of cosmic
  acceleration},'' {\em Physics Reports}~{\bf 530},  87--255 (Sept. 2013).

\bibitem{Bassett:2009mm}
Bassett, B.~A. and Hlozek, R.,  [{\em Dark
  Energy}{\nolinebreak\hspace{0.1em}]}, ch.~{Baryon Acoustic Oscillations},
  Cambridge University Press (2010).

\bibitem{Kogut:2007tq}
Kogut, A. et~al., ``{Three-Year Wilkinson Microwave Anisotropy Probe (WMAP)
  Observations: Foreground Polarization},'' {\em Astrophys. J.}~{\bf 665},
  355--362 (2007).

\bibitem{eisenstein05}
{}Eisenstein, D. J. e.~a., ``{Detection of the Baryon Acoustic Peak in the
  Large‐Scale Correlation Function of SDSS Luminous Red Galaxies},'' {\em The
  Astrophysical Journal}~{\bf 633},  560–574 (Nov 2005).

\bibitem{percival07}
{Percival, W.~J. et al.}, ``{Measuring the Baryon Acoustic Oscillation scale
  using the Sloan Digital Sky Survey and 2dF Galaxy Redshift Survey},'' {\em
  Monthly Notices of the Royal Astronomical Society}~{\bf 381},  1053–1066
  (Sep 2007).

\bibitem{abbott18}
Abbott, T.~M.~C. e.~a., ``{Dark Energy Survey Year 1 results: measurement of
  the baryon acoustic oscillation scale in the distribution of galaxies to
  redshift 1},'' {\em Monthly Notices of the Royal Astronomical Society}~{\bf
  483},  4866–4883 (Dec 2018).

\bibitem{Chang:2007xk}
Chang, T.-C., Pen, U.-L., Peterson, J.~B., and McDonald, P., ``{Baryon Acoustic
  Oscillation Intensity Mapping as a Test of Dark Energy},'' {\em Phys. Rev.
  Lett.}~{\bf 100},  091303 (2008).

\bibitem{chime19a}
{The CHIME/FRB Collaboration}, ``{Observations of fast radio bursts at
  frequencies down to 400 megahertz},'' {\em Nature}~{\bf 566},  230–234 (Jan
  2019).

\bibitem{chime19b}
{The CHIME/FRB Collaboration}, ``{A second source of repeating fast radio
  bursts},'' {\em Nature}~{\bf 566},  235–238 (Jan 2019).

\bibitem{chime19c}
{The CHIME/FRB Collaboration}, ``{CHIME/FRB Detection of Eight New Repeating
  Fast Radio Burst Sources},'' {\em The Astrophysical Journal Letters}~{\bf
  885}(1) (2019).

\bibitem{chime20a}
{Fonseca, E. et al.}, ``{Nine New Repeating Fast Radio Burst Sources from
  CHIME/FRB},'' {\em The Astrophysical Journal}~{\bf 891},  L6 (Feb 2020).

\bibitem{chime20b}
{The CHIME/FRB Collaboration}, ``{Periodic activity from a fast radio burst
  source},'' {\em Nature}~{\bf 582},  351–355 (Jun 2020).

\bibitem{swetz18}
{Swetz, D.~S. et al.}, ``{The Atacama Cosmology Telescope: The Receiver and
  Instrumentation},'' {\em The Astrophysical Journal Supplement Series}~{\bf
  194} (2018).

\bibitem{benson14}
{Benson, B.~A. et al.}, ``{SPT-3G: A Next-Generation Cosmic Microwave
  Background Polarization Experiment on the South Pole Telescope},'' {\em
  Society of Photo-Optical Instrumentation Engineers (SPIE) Conference Series}
  {\bf 9153} (July 2014).

\bibitem{levi13}
{Michael Levi and the DESI collaboration}, ``{The DESI Experiment, a whitepaper
  for Snowmass 2013},'' (2013).

\bibitem{brough20}
{Sarah Brough et al.}, ``{The Vera Rubin Observatory Legacy Survey of Space and
  Time and the Low Surface Brightness Universe},'' (2020).

\bibitem{gupta17}
{Gupta, N. et al.}, ``{The MeerKAT Absorption Line Survey (MALS)},'' (2017).

\bibitem{kuhn20}
{Kuhn et al.}, ``{Noise temperature testing for the Hydrogen Intensity and
  Real-time Analysis eXperiment (HIRAX)},'' {\em Society of Photo-Optical
  Instrumentation Engineers (SPIE) Conference Series} {\bf 11445} (Dec. 2020).

\bibitem{newburgh16}
{Newburgh, L.~B. et al.}, ``{HIRAX: A Probe of Dark Energy and Radio
  Transients},'' {\em Society of Photo-Optical Instrumentation Engineers (SPIE)
  Conference Series} {\bf 9906} (July 2016).

\bibitem{bandura14}
{Bandura, K. et al.}, ``Canadian hydrogen intensity mapping experiment (chime)
  pathfinder,'' {\em Society of Photo-Optical Instrumentation Engineers (SPIE)
  Conference Series}~{\bf 9145} (July 2014).

\bibitem{deboer17}
{DeBoer, David R. et al.}, ``{Hydrogen Epoch of Reionization Array (HERA)},''
  {\em Publications of the Astronomical Society of the Pacific}~{\bf 129},
  045001 (Mar 2017).

\bibitem{deng14}
Deng, M., Campbell-Wilson, D., and the CHIME~Collaboration, ``The cloverleaf
  antenna: A compact wide-bandwidth dual-polarization feed for chime,'' {\em
  International Symposium on Antenna Technology and Applied Electromagnetics
  (ANTEM)}~{\bf 16}, IEEE (July 2014).

\bibitem{oconnor20}
{O'Connor, P. et al.}, ``{The Baryon Mapping Experiment (BMX), a 21cm intensity
  mapping pathfinder},'' {\em Society of Photo-Optical Instrumentation
  Engineers (SPIE) Conference Series} {\bf 11445} (Dec. 2020).

\bibitem{king10}
{King, Oliver G. et al.}, ``{The C-Band All-Sky Survey: instrument design,
  status, and first-look data},'' {\em Society of Photo-Optical Instrumentation
  Engineers (SPIE) Conference Series}~{\bf 7741} (Jul 2010).

\bibitem{EwallWice:2016}
{Ewall-Wice}, A. et~al., ``{The Hydrogen Epoch of Reionization Array Dish. II.
  Characterization of Spectral Structure with Electromagnetic Simulations and
  Its Science Implications.},'' {\em The Astrophysical Journal}~{\bf 831},  196
  (Nov. 2016).

\bibitem{shahpari15}
Shahpari, M. and Thiel, D.~V., ``The impact of reduced conductivity on the
  performance of wire antennas,'' {\em IEEE Transactions on Antennas and
  Propagation}~{\bf 63},  4686–4692 (Nov 2015).

\bibitem{vanderlinde19}
{Vanderlinde, K. et al.}, ``{LRP 2020 Whitepaper: The Canadian Hydrogen
  Observatory and Radio-transient Detector (CHORD)},'' (2019).

\bibitem{shaw14}
{Shaw}, J.~R., {Sigurdson}, K., {Pen}, U.-L., {Stebbins}, A., and {Sitwell},
  M., ``{All-sky Interferometry with Spherical Harmonic Transit Telescopes},''
  {\em Astrophysical Journal}~{\bf 781},  57 (Feb. 2014).

\bibitem{shaw15}
{Shaw}, J.~R., {Sigurdson}, K., {Sitwell}, M., {Stebbins}, A., and {Pen},
  U.-L., ``{Coaxing cosmic 21 cm fluctuations from the polarized sky using m
  -mode analysis},'' {\em Phys. Rev. D}~{\bf 91},  083514 (Apr. 2015).

\end{thebibliography}
\bibliographystyle{spiebib} 

\end{document}